\documentclass[12pt,preprint]{aastex}
\usepackage{psfig}

\newcommand{\be}{\begin{equation}}
\newcommand{\en}{\end{equation}}
\def\ltsima{$\; \buildrel < \over \sim \;$}
\def\lsim{\lower.5ex\hbox{\ltsima}}
\def\gtsima{$\; \buildrel > \over \sim \;$}\def\gsim{\lower.5ex\hbox{\gtsima}}

\begin{document}

\title{The BMW Detection Algorithm applied to the Chandra Deep Field 
south: deeper and deeper}

\author{A.~Moretti\altaffilmark{1}, D.~Lazzati\altaffilmark{2},
S.~Campana\altaffilmark{1}, G.~Tagliaferri\altaffilmark{1}}

\altaffiltext{1}{Osservatorio Astronomico di Brera, Via E. Bianchi
46, Merate (LC), 23807, Italy.}
\altaffiltext{2}{Institute of Astronomy, University of Cambridge, 
Madingley Road, Cambridge CB3 0HA, UK.}

\authoremail{moretti@merate.mi.astro.it}

\begin{abstract}
Chandra deep fields represent the deepest look at the X--ray sky. 
We analyzed the Chandra Deep Field South (CDFS) with the aid of a dedicated 
wavelet-based algorithm. Here we present a detailed description of the 
procedures used to analyze this field, tested and verified by means of
extensive simulations. 
We show that we can safely reconstruct the Log N--Log S source distribution 
of the CDFS down to limiting fluxes of $2.4\times 10^{-17}$ and $2.1\times 
10^{-16}$ erg s$^{-1}$ cm$^{-2}$ in the soft (0.5--2 keV) and hard (2--10 keV) 
bands, respectively, fainter by a factor $\gsim 2$ than current estimates. At
these levels we can account for $\gsim90\%$ of the 1--2 keV and 2--10 keV
X--ray background.   
\end{abstract}

\keywords{X--ray background -- number-counts -- detection algorithm}

\section{Introduction}

The Chandra observatory is providing the astronomical community with
the deepest X--ray look at the sky (Mushotzky et al. 2000; Hornschemeier 
et al. 2000; Giacconi et al. 2001). Two 1 Ms observations have been
recently carried out: one in the northern hemisphere on the Hubble Deep
Field North (HDFN) and the other in the southern hemisphere on a field, named 
Chandra Deep Field South (CDFS), selected for its low column density
(the southern twin of the Lockman hole) and for the lack of bright
X--ray and optical sources.
The data are available from the Chandra public archive\\ 
({\tt http://asc.harvard.edu/udocs/ao2-cdf-download.html})
A further 1 Ms data set on the HDFN will be available in the next future.

The main goal of these observations is to look at the X--ray sky at
the deepest level and to gain insight in the population of emitting
sources comprising the cosmic X--ray background (XRB). 
The details of the data reduction and analysis procedures
that have been applied to manage the CDFS data-set are discussed in
Section 2. In order to fully exploit the potential of these data, refined
detection algorithm have to be used. We developed a wavelet-based source
detection algorithm (Lazzati et al. 1999; for its main characteristics see
also Section 3), that we applied to the full sample of ROSAT HRI fields 
(Campana et al. 1999; Panzera et al. 2002, in preparation). We modified this
detection  
algorithm, called Brera Multi-scale Wavelet (BMW), to account for the
specific characteristics of the Chandra ACIS Imaging and Spectroscopic
instruments. The algorithm (BMW-Chandra) has been
extensively tested in the extreme conditions provided by the CDFS
(Section 4). Our final goal is to obtain a source detection and a
source Log N--Log S at the faintest limits in order to resolve as much as
possible of the XRB in point sources. In order to compare
our results with previous investigations, we carry out the analysis in the 
soft (0.5--2 keV) band and in the hard (2--10 keV) band. We are able
to safely reconstruct the Log N--Log S source flux distribution down to
$2.4\times 10^{-17}$ and $2.1\times 10^{-16}$ 
erg s$^{-1}$  cm$^{-2}$ in the soft and hard bands, respectively (Section 5). 
A first account of these results have been given in Campana et al. (2001), 
here we extend these results discussing in more details the resolved
background and source characteristics. Conclusions are reported in Section 6. 


\section{The data}
 
All exposures were taken with the Chandra X--ray Observatory (Weisskopf et
al. 2000) Advanced CCD Imaging Spectrometer (ACIS-I) detector (Bautz et
al. 1998). ACIS-I consists of four CCDs arranged in a $2\times 2$ array 
assembled in an inverse shallow pyramid configuration to better follow 
the curved focal surface of the mirrors. The full ACIS-I has a field of 
view of $16.9'\times 16.9'$. The on-axis image quality is $\sim 0.5''$ FWHM 
increasing to $\sim 3.0''$ FWHM at $\sim 4.0'$ off-axis. 
The CDFS has been obtained performing eleven exposures of the same
area of the sky with slight position and orientation offsets. A total
exposure time of 966 ks is obtained by summing them together. In
Tab. \ref{tab:subexp} we report the main properties of each
exposure. 

First we have to match the position of the different exposures to
add them into a single image. To this aim we performed a preliminary
source detection on the 11 sub-frames at full resolution.
We used the grid of the positions of the brightest sources to calculate 
the ten roto--translation matrices with respect to the first image.
In all ten matrices we find that the rotation terms are negligible.
The translation factors of each frame with their uncertainties are
reported in Tab. \ref{tab:subexp}. The reported uncertainties are the
standard deviations of the position distributions of the grid sources
after the translations.

Starting with the level 2 event files, the data were filtered to include
only the standard event grades (corresponding to the ASCA grades 0, 2, 3, 4,
6). In order to compare our
results with previous studies and existing estimates of the cosmic
X--ray background, we selected a 0.5--2 keV soft energy band and a 2--7
keV hard energy band to carry out the scientific analysis. The limit 
at 7 keV has been selected because at higher energies the effective 
area decreases and the background increases, resulting in a lower signal 
to noise of celestial sources. The fluxes are then extrapolated to the 
2--10 keV energy band.

Chandra observations are affected by the `space weather',
resulting in periods with high background. In order to eliminate these
periods we calculated the total counts light curve for each
exposure using a time resolution such that in each bin we had $\sim 400$
events (i.e. $\sim 1,000$~s in the soft band and $\sim 600$ s in the hard
band). We excluded all time 
intervals above $3\,\sigma$ of the mean. From a total
of 966 ks, we excluded 24 ks in the soft band (28 ks in
the hard band), giving a final effective exposure time of
942 ks (939 ks in the hard). The lost fractions amount to $\sim 3\%$
of the total time, but allowed us to reduce considerably the background 
in each band.
In particular, we registered a very strong flare in the observation
ID1431 which contained about $90\%$ of the total events
both in the soft and hard band in about $10\%$ of the exposure time.
 
CCDs are usually affected by defects such as bad columns, bad and
hot/flickering pixels and cosmic ray events. The great majority of
these defects are already eliminated by the standard Chandra
processing pipeline. Random flickering pixels (probably excited by
energetic cosmic rays) however do occur and have to be removed manually.  
These are active pixels for only 2--6 frames and might be
easily detected as X--ray sources. Following Tozzi et al. (2001), we
defined a pixel as ``flickering'' when it registers two events within
the time of two consecutive frames and in each observation we
eliminated all the events registered from that pixel (this can be done safely 
thanks to the lack of bright sources). In this way we
excluded about 600 events in the soft band whereas in the hard band we did not
find any flickering pixels. We also eliminated the data from chip 3 of the ID581
observation, because it suffered from Good Time Intervals inconsistences
with other chips, resulting in an anomalous high background level.

The count-rate to flux conversion factors in the 0.5--2 keV and in the 
2--10 keV bands were taken from Tozzi et al. (2001). These were computed 
using the response matrices at the aim point and amount to $(4.6\pm 0.1) 
\times 10^{-12}$ erg s$^{-1}$ cm$^{-2}$ per count s$^{-1}$ (0.5--2 keV), and 
$(2.9\pm 0.3) \times 10^{-11}$ erg s$^{-1}$ cm$^{-2}$ (2--10 keV) per count 
s$^{-1}$ in the hard band, assuming a Galactic absorbing column of 
$8\times 10^{19}$ cm$^{-2}$ and a photon index $\Gamma = 1.4$, i.e. the
spectrum of the XRB. Flux uncertainties are derived considering 
power law indices in the range $\Gamma=1.1-1.7$. Fluxes are corrected for
vignetting by using the exposure map (see below).

The analysis has been carried out on images rebinned by a factor of 2
(i.e. 1 pixel corresponds to 0.98 arcsec), allowing us to deal
directly with the entire image. We found that working on sub-images
at the natural scale improves slightly the positional accuracy at the
price of a four times longer computational time and with the problem of
matching the four sub-images.  

In the analysis of the CDFS we have
restricted our analysis to the central 8 arcmin radius circle,
assuming as the center the aim point of ID581 observation. This
is a good compromise between the simplicity of the geometry
and the efficiency of the data analysis (see below).
The minimum value of the exposure map
in this region is about $23\%$ of the maximum value. This occurs at
the border of the circle where the exposure map has a very steep
decrease. More than $90\%$ of the exposure map within the 8
arcmin central circle have values larger than $80\%$ of the maximum
value, corresponding to an effective exposure time greater then
753~ks.
The average background levels in this region are 0.18 and
0.29 counts per binned pixel in the soft and hard band, respectively. These
correspond to 0.07 and 0.11 counts s$^{-1}$ per chip and are in very good
agreement with the expected value reported in the Chandra Observatory Guide.

\section{The BMW algorithm}

The wavelet transform (WT) is a mathematical tool able of
decomposing an image in a set of sub-images, each of them carrying the
information of the original image at a given scale. These features
make the WT well suited for the analysis of X--ray images, where the scale
of sources is not constant over the field of view. In addition, the
background is automatically subtracted since it is not characterized
by any scale. The use of WT as X--ray detection algorithms was
pioneered by Rosati et al. (1995; 1998) for the detection of extended
sources in ROSAT PSPC fields and subsequently
adopted by many groups (Grebenev et al. 1995; Damiani et al. 1997a;
Pislar et al. 1997; Vikhlinin et al. 1998; Lazzati et al. 1998; Freeman et
al. 2001). 

The BMW detection algorithm is a WT-based
algorithm for the automatic detection and characterization of sources
in X--ray images. It is fully described in Lazzati et al. (1999) and
has been developed to analyze ROSAT HRI images producing the BMW-HRI 
catalog (Campana et al. 1999; Panzera et al. 2002).  
We have recently updated this algorithm to support the analysis
of Chandra ACIS Imaging and Spectroscopic images. Here we
summarize the basic steps of the algorithm. First, the WT of the input
image is performed.  The BMW WT computation is based on the discrete
multi-resolution theory and on the ``\'a trous'' algorithm (Bijaoui
et al. 1991). This is different with respect to
algorithms based on a continuous WT, which can sample more scales at
the cost of longer computing time (Rosati et al. 1995; Grebenev et
al. 1995; Damiani et al. 1997a). We used a Mexican hat mother, which can
be analytically approximated by the difference of two Gaussians
(Slezak et al.  1994).  We performed the Chandra ACIS-I data analysis
with a rebin factor of 2 and we used the scales $a=1$, 2, 4, 8, 16 and 
32 pixels, where $a$ is the scale of the transform (see Lazzati et al. 1999).

Candidate sources are identified as local maxima above the threshold
in the wavelet space at each scale.  A catalog for each scale is
obtained.  Different scale catalogs are then cross--correlated and
merged in a single catalog. For each scale the threshold in the
wavelet space is calculated by means of Monte Carlo simulations: for a
grid of values of Poissonian backgrounds in the range between
$10^{-4}- 10^1$ counts per pixel we estimated the number of
expected spurious detections as a function of the threshold value.
The background value is measured as an average on the whole image by
means of a $\sigma$-clipping algorithm. The number of spurious sources
per field can be fixed arbitrarily by the user (typical values are in
the range $0.1-10$).  At the end of this detection step we have a
first estimate of the source position, an estimate of the counts from
the local maximum value of the WT and a guess of the size from the
scale where the WT is maximized.

The final catalog is obtained through the characterization of the
sources, which is performed by means of a $\chi^2$ minimization with
respect to a Gaussian model source in the wavelet space.  The WT of a
detected source is used at three different scales, with the central
being the scale at which the source has been found (corrections are
introduced when the best scale is the first or the last).  In order to fit
the model on a set of independent data, the WT coefficients are decimated
according to the scheme described in Lazzati et al. (1999).
Neighbor sources can be fitted simultaneously in the deblending process, 
allowing the characterization of faint objects located near bright sources.

\subsection{Total background map}

\begin{figure*} [bht]
\begin{tabular}{cc}
{\psfig{figure=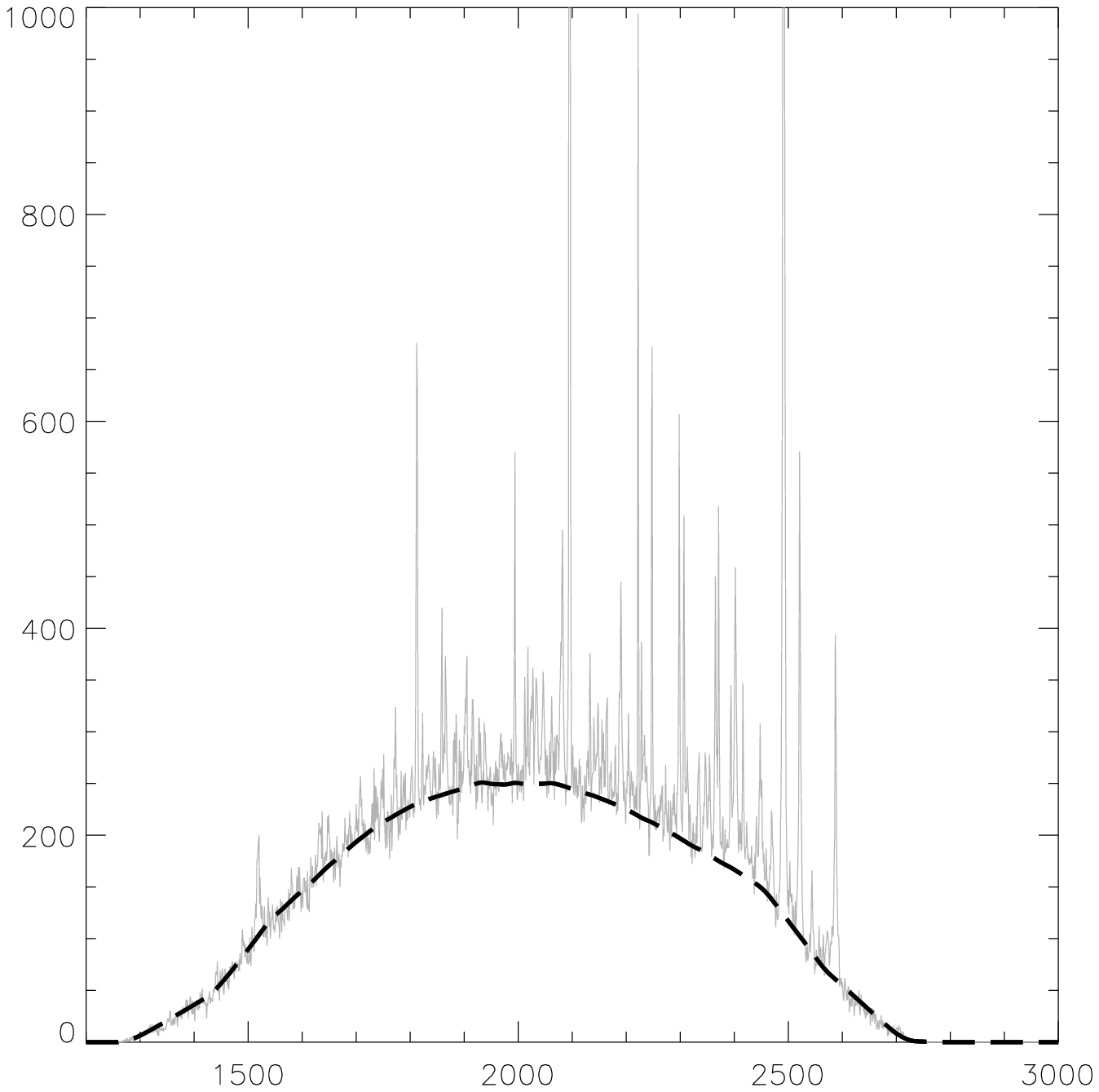,width=8cm}}&{\psfig{figure=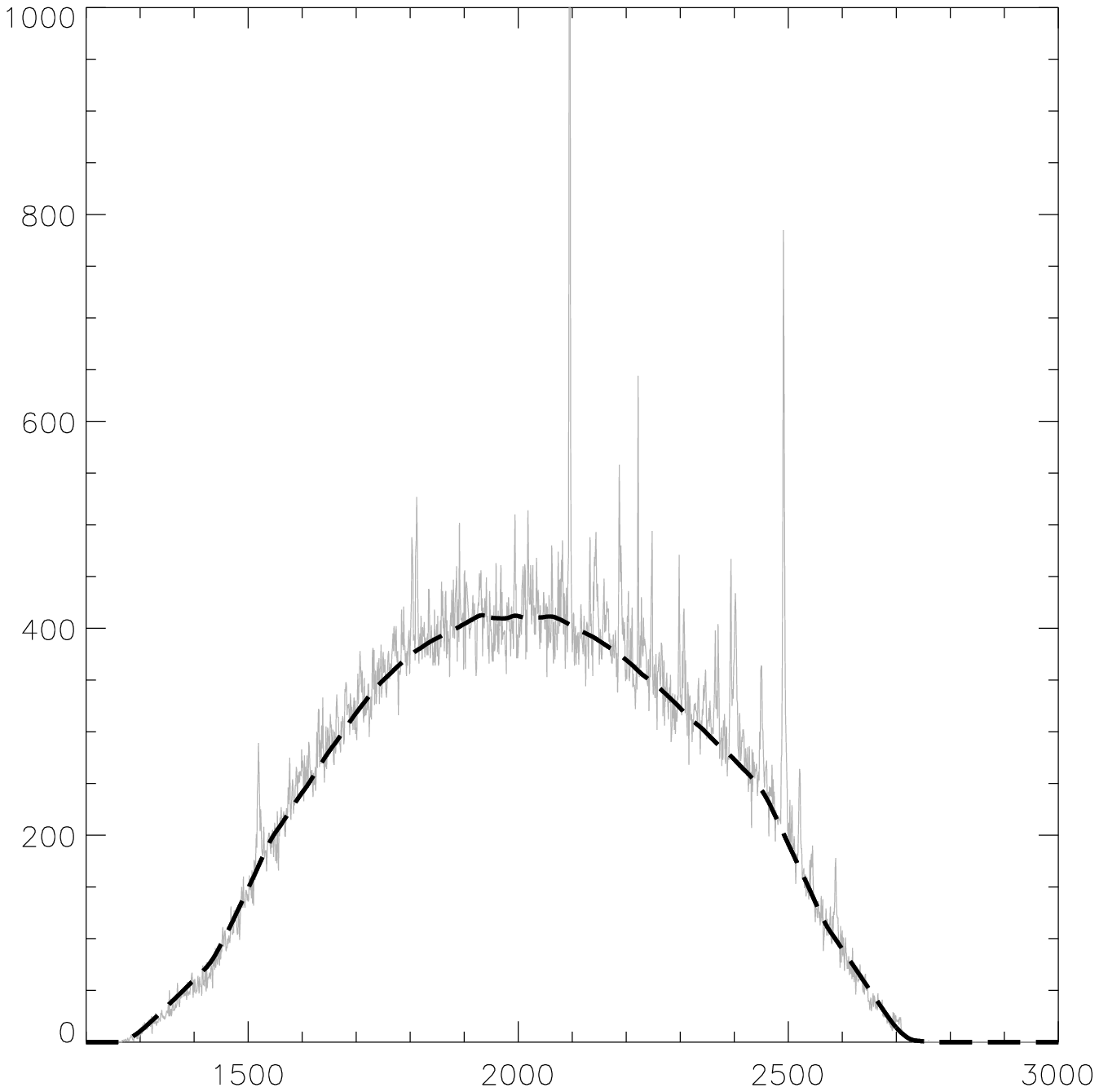,width=8cm}}\\
\end{tabular} 
\caption[]{The comparison between the total background maps and the real data in 
the soft band (left panel) and hard band (right panel). We compare the 
projections of the columns of the accumulated images (grey solid line)
with the projection of the total background map
(black dashed line). On the X axis the units are rebinned pixels
and on the Y axis are simply the counts.}
\label{fig:figmap} 
\end{figure*}

The WT detection is carried out on top of a map representing at best the image
background (named total background map, see Lazzati et al. 1999).  
The ACIS-I background consists of two different components, the
cosmic X--ray background and cosmic ray-induced events (also called
particle background).
The first is made by focalized X--ray photons from not resolved sources and
therefore suffers from mirror vignetting.
Its spectrum can be modelled as a power law with photon index $\Gamma = 1.4$.
Following the CIAO 2.1 Science Threads
({\tt http://asc.harvard.edu/ciao/threads}) we built the exposure maps 
for each of the eleven observations (both in the soft and in the hard
band) assuming a photon index $\Gamma = 1.4$ as input spectrum.
The second contribution has a spatial pattern which is basically due to the
electronics of the system and depends on its temperature. 
Based on the Chandra ACIS background web page\\
({\tt http://asc.harvard.edu/cal/Links/Acis/acis/Cal\_prods/bkgrnd/current/index.html})
we found that for the ACIS-I chips in the energy range and in
the spatial region of interest for the only particle background 
we can assume a spatial flat pattern with an inaccuracy of less than $10\%$.
We assumed that the particle background in the soft band is $50\%$ of the
total and the $66\%$ in the hard band with a flat energy spectrum (Baganoff 1999).
Summing the two components of the background we obtain a total background map
and taking a Poissonian realization of it we can accurately account for the data
in both bands as shown in Fig. \ref{fig:figmap}. 
This procedure is slightly different from the one adopted in Campana et
al. (2001), where a map based only on the cosmic X--ray background was
considered.  


\section{Simulations}

We tested the detection algorithm, trying to reproduce the
real data in terms of background level and source flux distribution,
matching the characteristics of the CDFS fields. The analysis
is carried out in two energy bands: the soft (0.5--2 keV) and the hard
(2--7 keV) band. 
 From the detection algorithm point of view the two bands differ for 
two main reasons: the shape of the exposure maps and the different background
level. This forced us to carry out two different sets of simulations.

X--ray sources were generated at random positions in a $30' \times 30'$
square region around the center of the
image (larger than the real field of view) and with fluxes distributed
in the range $10^{-17}-10^{-13}$ erg s$^{-1}$ cm$^{-2}$ for the soft
band and $10^{-16}-10^{-13}$ erg s$^{-1}$ cm$^{-2}$ in the hard band.
The number counts have been generated according to the Log N--Log S
integral distributions reported in Tozzi et al. (2001), i.e. power laws with
slope of $0.66$ in the soft band and $0.92$ in the hard band.
 
The CDFS is made of eleven observations. For this reason
we added a further element to our simulations: for each source 
we calculated its position in each observation and derived the corresponding 
off-axis angle (each source has different positions in the detector in 
each different observation, depending on the coordinate of the pointing, 
which are slightly different, and on the roll angle).  Then we calculated 
the expected number of photons for each observation. These depend on 
the conversion factor and on the local value of the exposure map.  Finally,
for each single observation we 
multiplied the expected number of photons with the corresponding point
spread function image (from the Chandra {\tt CALDB}) and we summed the
eleven images to the background image. This
procedure is aimed at reproducing at best the image characteristics.

We used these  routines instead of the MARX simulator essentially because of
computational reasons: we found that dealing with the input/output of
data was easier for us using our IDL routines.  We repeated the procedure
for 100 times, for a total sample of about 150,000 input sources (here
and in the following we refer to simulations in the soft band, unless
otherwise stated; in the hard band results are similar). We recovered
25,000 of them within the central 8 arcmin radius, from which we were 
able to verify the performance of the procedure.

\subsection{Spurious detection and position determination}

\begin{figure*} [bht]
\begin{tabular}{cc}
{\psfig{figure=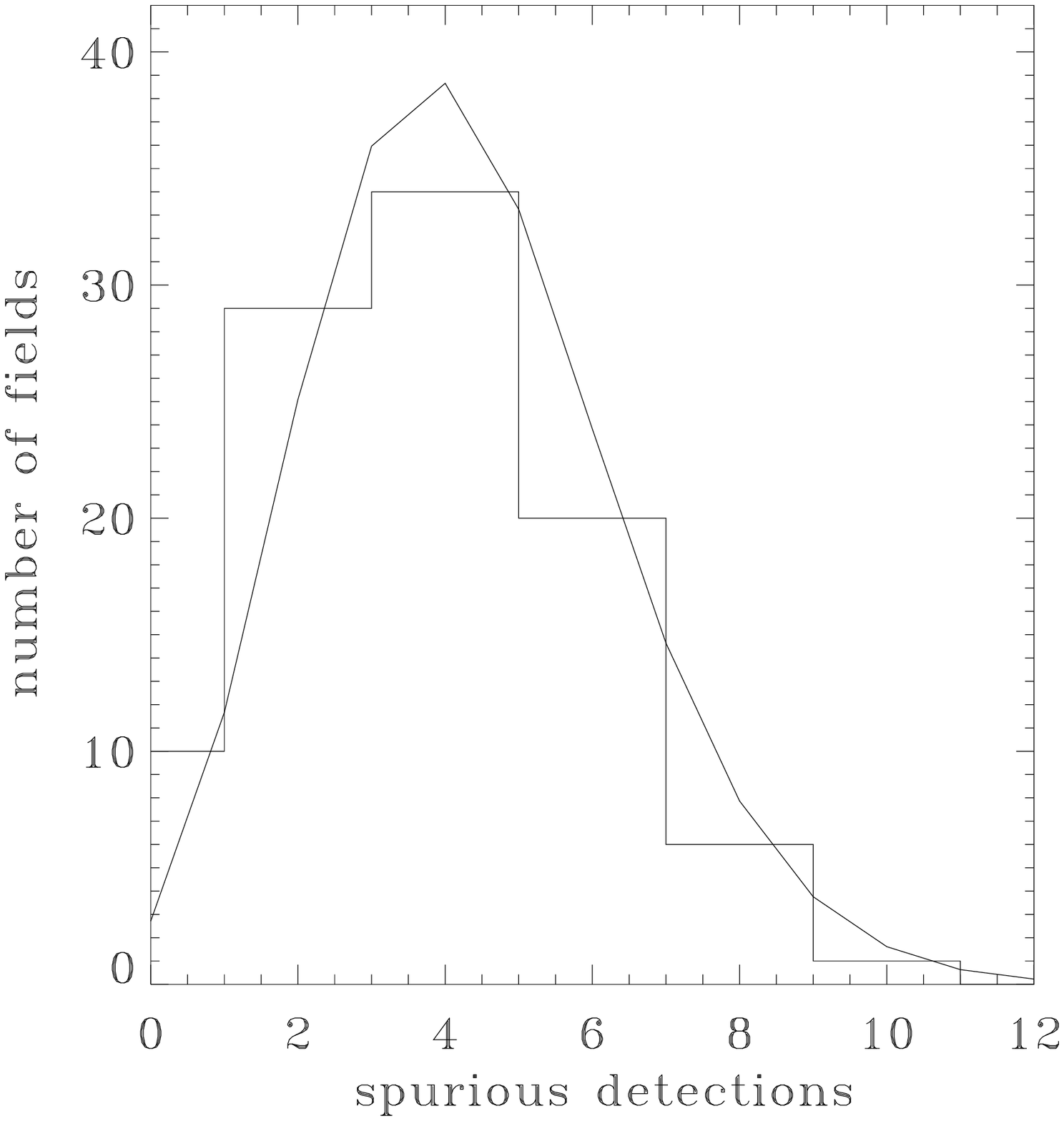,width=8cm}}&{\psfig{figure=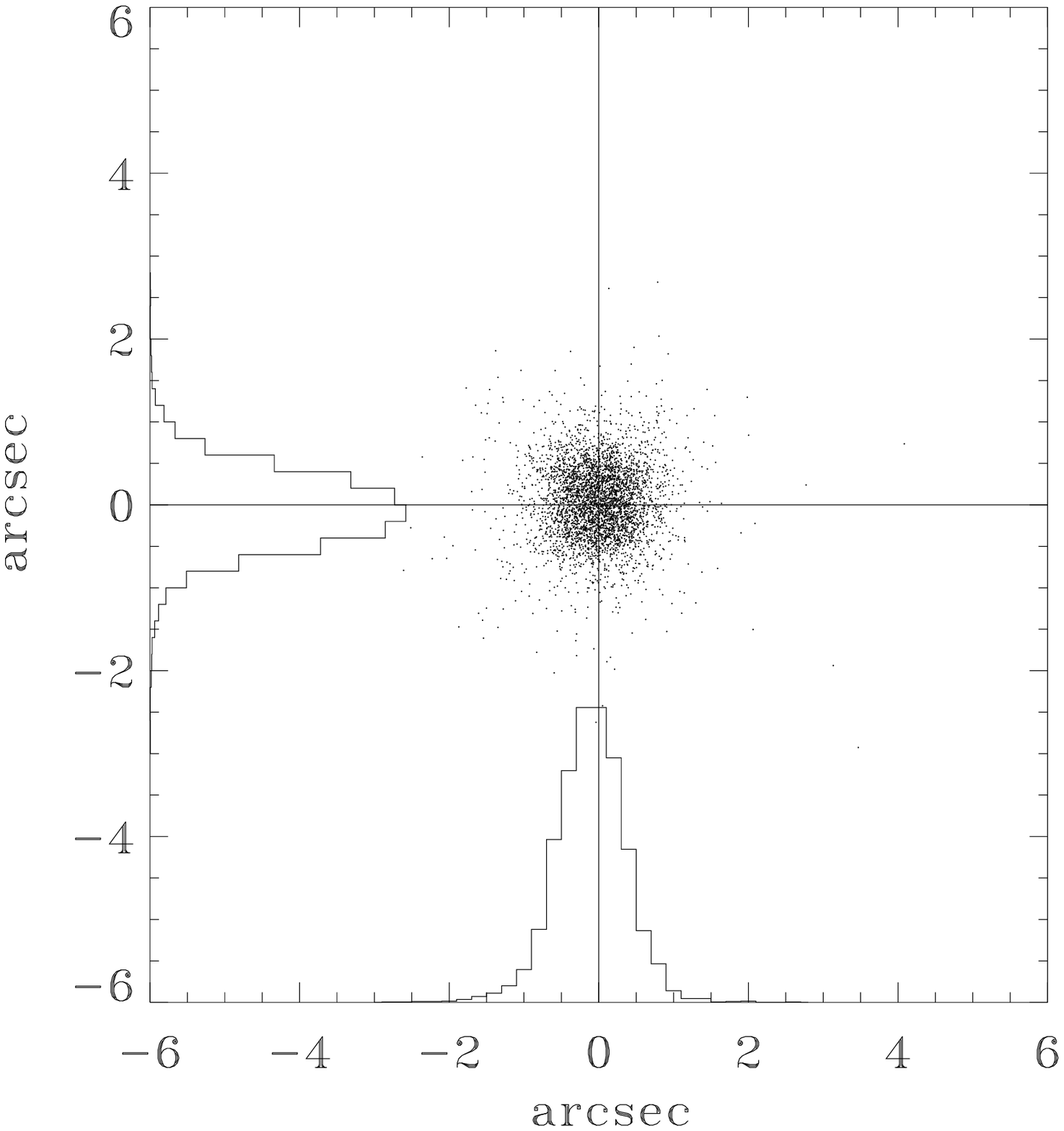,width=8cm}}\\
\end{tabular} 
\caption[]{The performance of the BMW detection algorithm.
In the left panel, the distribution of the of number of spurious
detections in the simulated fields is plotted and compared with the
expected Poissonian distribution with a mean value of 4.3. In the right
panel we plot the differences between the input position of the
simulated sources across the $8'$ radius field of view and the
position recovered by the detection algorithm. The difference 
distributions are very well approximated by Gaussians centered in
zero and with a standard deviation of 0.6 arcsec.}
\label{fig:spuepos} 
\end{figure*}

In the BMW detection algorithm the expected number of spurious
sources is one of the fundamental input parameters. 
In our analysis we fixed the number of spurious sources to a total of
4.3 in the 8 arcmin radius circle (which corresponds to 6 expected spurious
in a $1024\times1024$ pixels image).
In 100 simulated frames we found a total of 425 spurious sources 
(431 expected, see left panel of Fig. \ref{fig:spuepos}).
This shows that the contamination of our sample is well understood and
under control. This corresponds to sources detected at a significance level
larger than $4\,\sigma$. 

Source positions are recovered accurately. In the right panel of
Fig. \ref{fig:spuepos} we plot the differences between the input and
output position of simulated sources. The difference distributions 
on both axes are well approximated by Gaussian functions
centered on zero and with a r.m.s. of $0.6''$ (i.e. $\sim$ half a pixel
at rebin 2, at which the analysis has been carried out).

\subsection{Sky--coverage}

The distribution of the detected sources is usually depleted at the faint 
end. In order to compute the Log N--Log S it must be corrected 
for (i) detection probability lower than one close to the
sensitivity limit (completeness function; CF) and (ii) the
different areas effectively surveyed at different fluxes (sky
coverage; SC).
The CF is due to the fact that sources are preferentially detected if 
they sit on a positive background fluctuation, while they are missed 
in the opposite case. The proper SC takes into account that the sensitivity 
limit can change in the same observation if the background or
spatial resolution are not uniform or in different observations when different 
exposure times are considered.
In single observations the SC depends on the off--axis angle, since
the detector sensitivity decreases (lower effective exposure time) and
the spatial resolution worsen.
In the CDFS the geometry is complicated by the sum of observations
with slightly different aim points and different roll angles.
We have verified that this fact does not affect significantly the symmetry
of the SC. 
More importantly, the SC has strong variations 
in those locations of the sky that were not imaged in all the eleven
observations. In order to avoid large SC fluctuations, we restricted our 
analysis to the inner $8'$ of the field, where the background and exposure 
are roughly constant (see above).

\begin{figure*} [hbt]
\begin{tabular}{cc}
{\psfig{figure=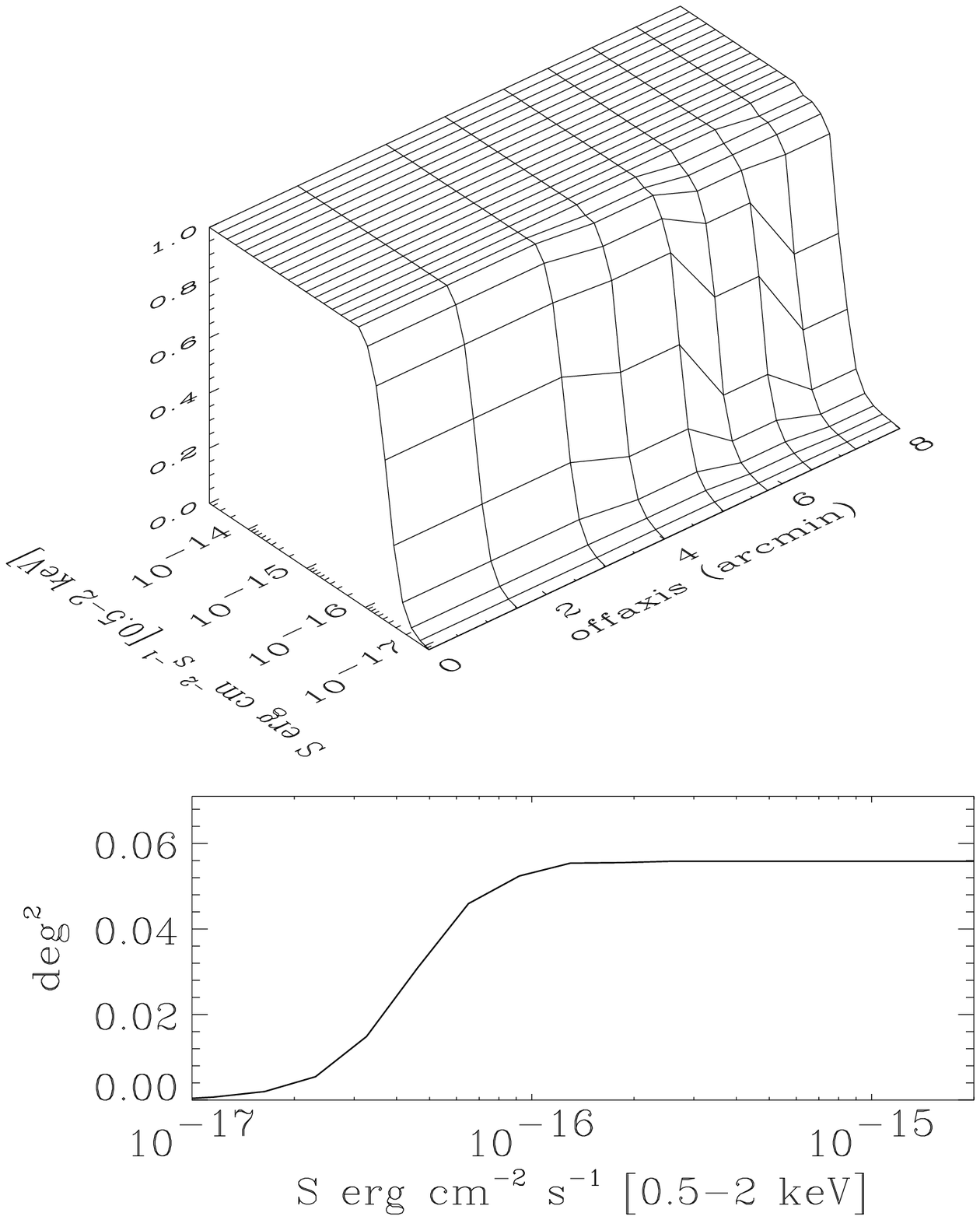,width=8cm}}&{\psfig{figure=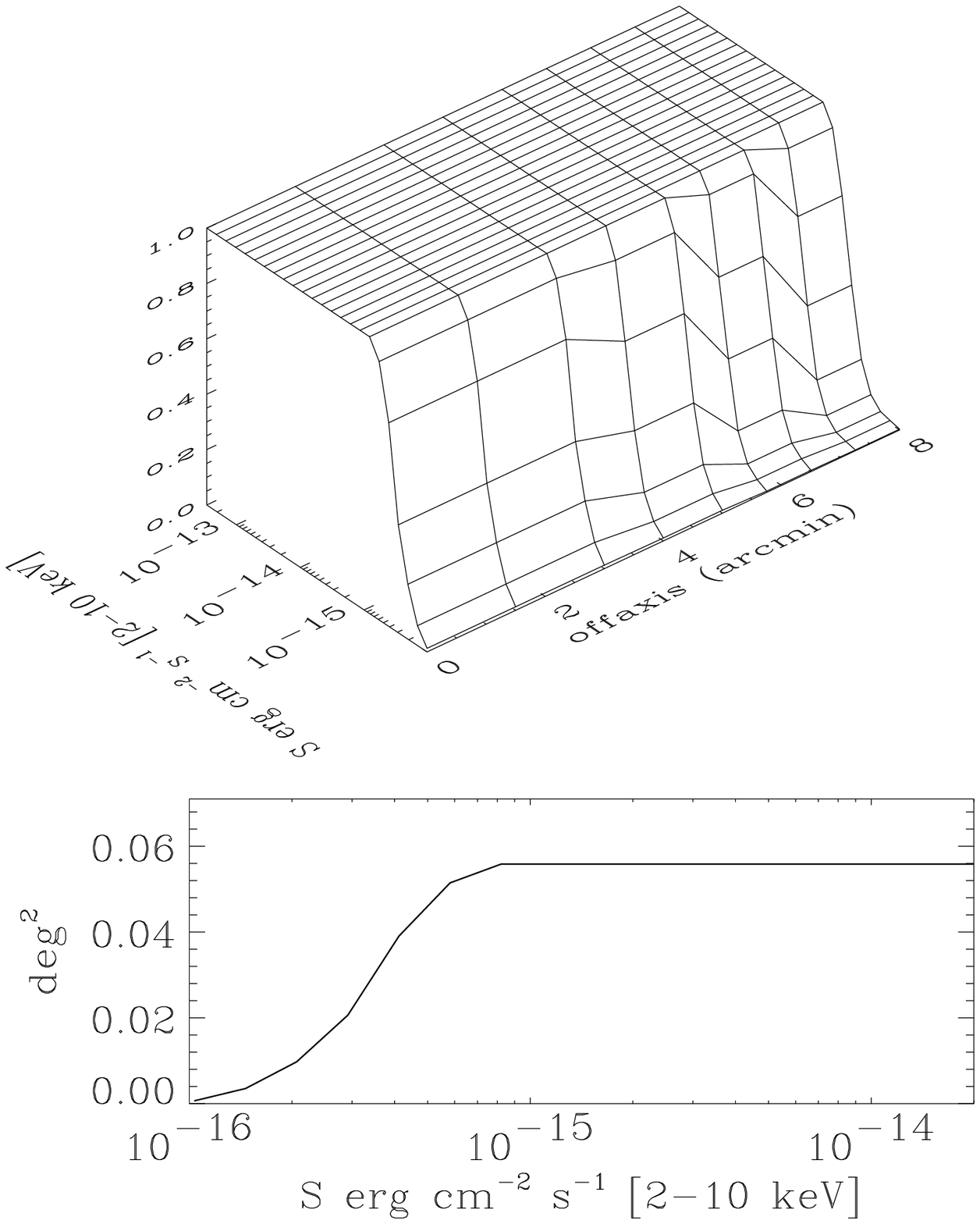,width=8cm}}\\
\end{tabular} 
\caption[]{In the upper part of the figure the graphics of the probability
of detecting a source (completeness function) in the soft and hard bands are shown.
They are functions of the flux of the sources and due to the combination 
of the PSF broadening and the vignetting they are also functions of the
distance from the center of the image (i.e. the off-axis angle).
We have divided the CDFS in 4 circular coronae and calculated 
4 different completeness functions to take into account this effect.
In the lower panels the total sky coverages in the soft and hard band in unit
of square degrees are plotted: they are calculated as the sum of the
contributions from the 4 different circular coronae.}
\label{fig:skycov} 
\end{figure*}

Using WT based detection algorithms the SC can be computed either
analytically or by means of simulations. The first approach is
recommended when the intrinsic size of sources is poorly known (like
in surveys of cluster of galaxies, see Rosati et al. 1995). In the CDFS
the presence of extended sources is a minor effect for the computation
of the Log N--Log S. For this reason we computed the SC by comparing the
input and output number counts of simulated fields. We divided the 
selected field of view into four circular regions ($0'-3'$, $3'-5'$,
$5'-6.5'$ and $6.5'-8'$) and derived the SC and CF from
simulations as a single correction (which we hereafter will call sky
coverage). In the upper panels of Fig. \ref{fig:skycov}
the derived CF as a function of flux and off-axis angle is shown 
for the soft and hard band; in the lower panels the total sky-coverages
are shown in unit of square degrees.

\subsection{Fluxes and bias correction}

\begin{figure*} [hbt]
\begin{tabular}{cc}
{\psfig{figure=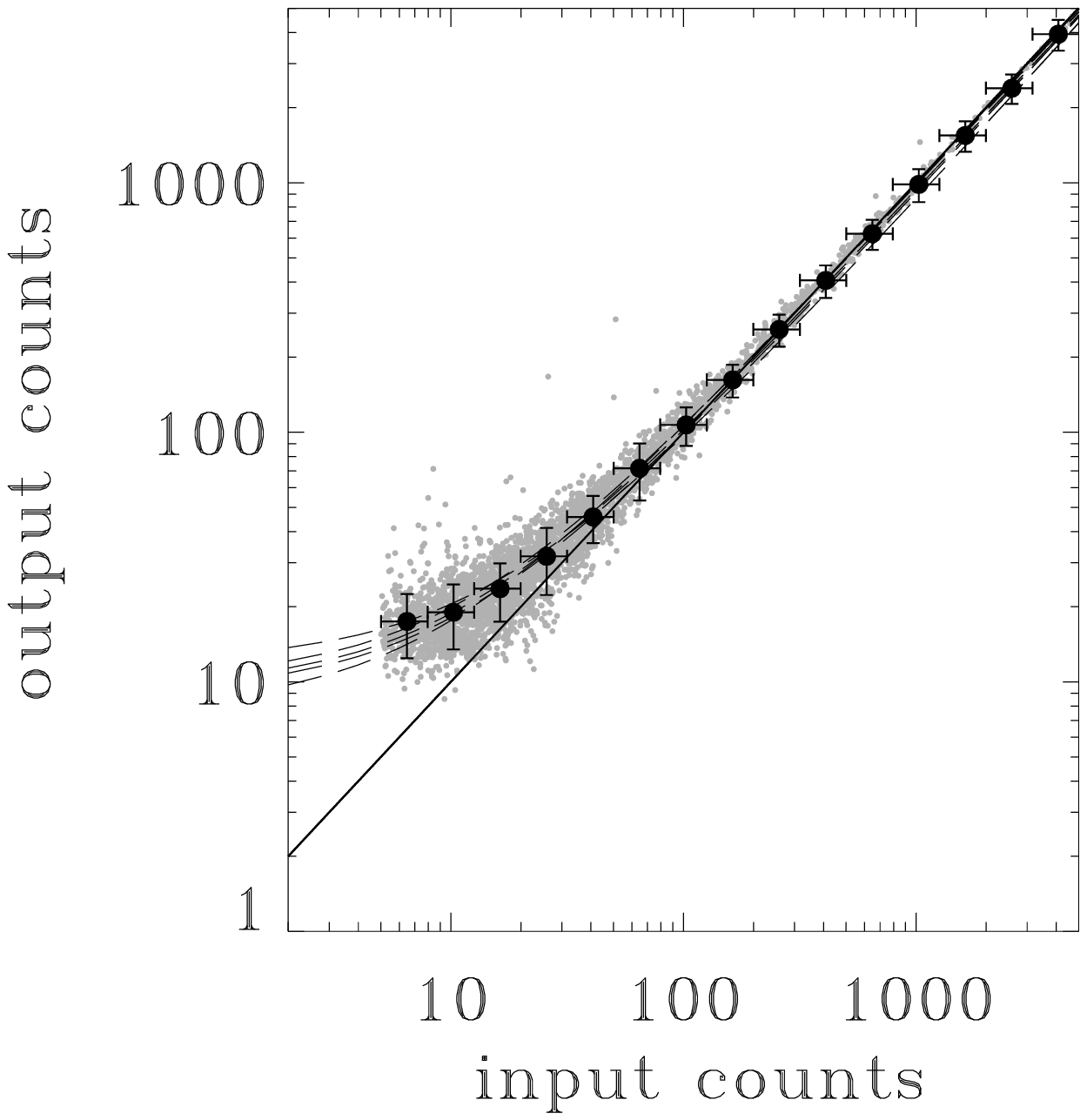,width=8cm}}&{\psfig{figure=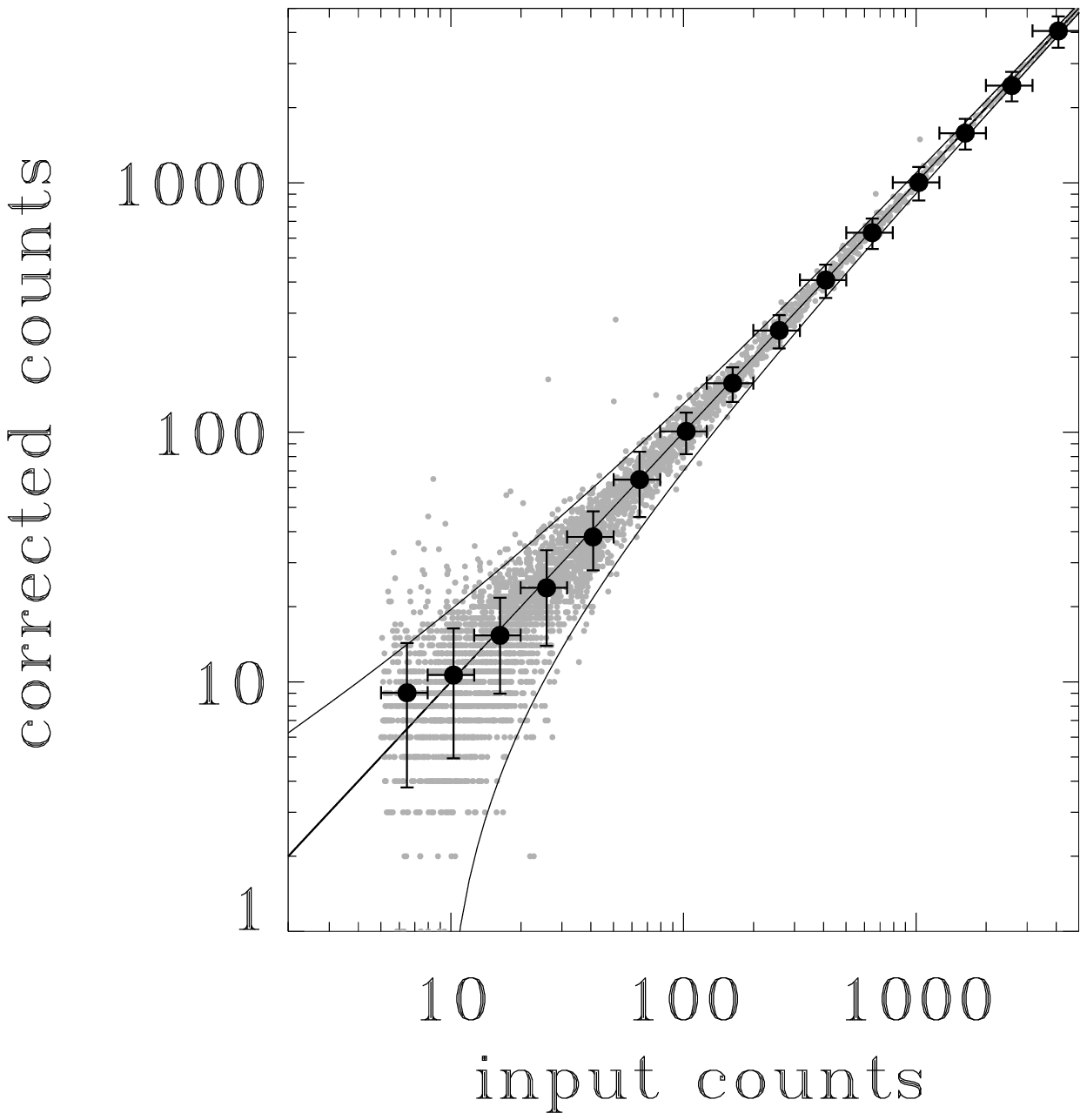,width=8cm}}\\
\end{tabular} 
\caption[]{In the left panel the measured counts versus input simulated counts
in the soft band are plotted (little grey dots). It is evident that the
detection limit is about 10 counts and where the detection is incomplete
(below 30 counts) the measured counts are overestimated. The dashed lines show 
the analytical approximation of the Eddington bias in the measure of faint
fluxes for different off-axis angles. The solid line is the straight lines $ y=x $.
The black big dots are the mean values of the output counts for different
bins of input counts.
The right panel shows the counts corrected by the inverse of the analytical
function (little grey dots). The black big dots are the mean values of the
corrected counts for different bins of input counts.
Overplotted with the the solid lines are the straight lines $ y=x $ and 
the $90 \%$  expected Poisson uncertainties.}

\label{fig:fluxcor} 
\end{figure*}

In Fig. \ref{fig:fluxcor} we show the comparison between the input and
output counts of simulated sources (output fluxes are corrected for the 
flux lost during the WT characterization, i.e. the so-called Point Spread
Function (PSF) correction, evaluated on bright isolated sources for every
arcmin of the field of view, and vignetting, evaluated directly from the
exposure map). For high signal to noise sources the
counts are measured correctly, while at lower signal to noise ratio the count
distribution of the detected sources suffers from the well known
Eddington bias (e.g. Hasinger et al. 1993). In our case  
the bias starts affecting the data below $\sim 30$ input counts
in the soft band (corresponding to $\sim 10^{-16}$ erg s$^{-1}$ cm$^{-2}$). 
This bias is due to the superposition of faint sources (below the detection
threshold) on positive background fluctuations, making them detectable. 
Given the background level we expect to detect sources with less than 
10 counts only in correspondence of background peaks (or when two faint 
sources are merged together into a brighter one).  
It follows that the counts of the very faint sources are necessarily 
overestimated (otherwise they could not be detected).
We corrected for this bias following the approach of Vikhlinin et
al. (1995).  We fitted the output source counts as a function of the
input counts with polynomial functions for a grid of circular coronae,
since the bias is a function of the source width (see Fig. \ref{fig:fluxcor}).
By inverting the fitting functions, we get an unbiased estimate of the
source counts in the faint tail of the simulated sample. Input counts
versus corrected output counts are shown in the right panel of
Fig. \ref{fig:fluxcor}.

\begin{figure*} [hbt]
\begin{tabular}{cc}
{\psfig{figure=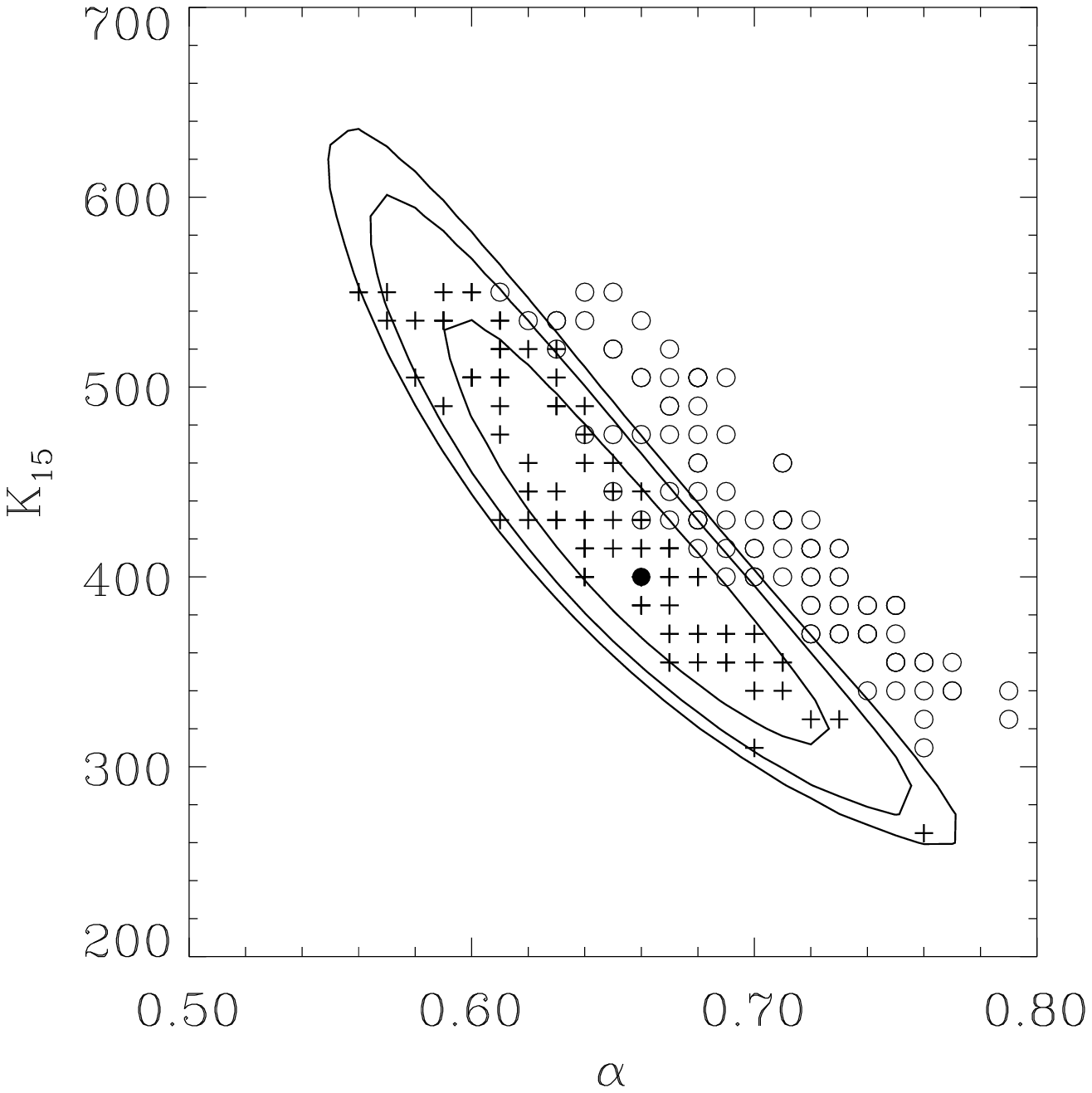,width=8cm}}&{\psfig{figure=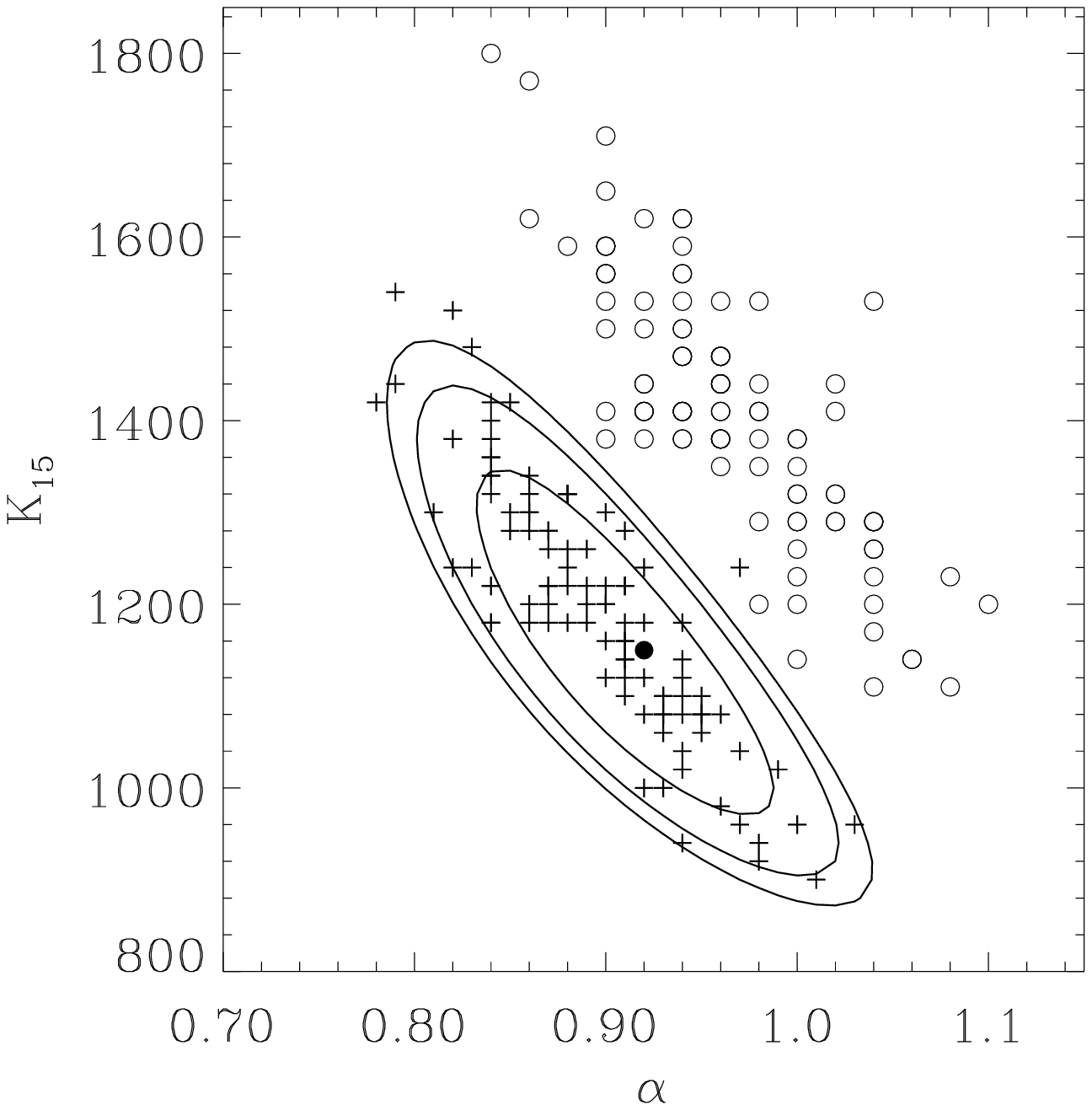,width=8cm}}\\
\end{tabular} 
\caption[]{
Left panel: values of the normalization $K_{15}$ (in units of 
$2\times 10^{-15}$ erg s$^{-1}$ cm$^{-2}$) and the slope $\alpha$ of the
source integral distribution as obtained for each of 100 simulated
fields. Simulations are carried in the soft 0.5--2 keV energy band.
Crosses refer to fits with the Eddington bias correction, whereas open circles
to fits on uncorrected data. The central filled dot indicate the input values 
($K_{15}=400$; $\alpha=0.66$).
The $68\%$, $90\%$ and $99\%$ confidence level are overplotted. 
Right panel: same as in the left panel but for the hard 2--10 keV energy band.
Input values are $K_{15}=1150$; $\alpha=0.92$.
}
\label{fig:100f_s} 
\end{figure*}

To assess the goodness of our correction procedure we perform a maximum 
likelihood fit to the differential (unbinned) flux distributions of the 
100 simulated fields individually and we compare the results of the fits 
with the expected values from the input distributions. In
Fig. \ref{fig:100f_s} we compare the results before and after the bias
correction. As explained below, in the final test section, for the corrected 
fluxes we set the flux limit at 5 and 7 counts in the innermost region 
in the soft and hard band, respectively, whereas for the uncorrected fluxes 
we are forced to use limiting flux of 10 and 12 counts, respectively. From the 
results of the maximum likelihood fit it is evident that after the flux
correction results are in very good agreement with the input value, 
whereas the distribution of the uncorrected fluxes suffers for a bias. 
This effect is more evident in the hard band where the background level 
is higher.  

A possible alternative to the correction of the measured fluxes is the
use of a completeness function allowing for a probability higher than
unity for those fluxes where faint sources are wrongly detected. Such
a probability function can be obtained by comparing the number of
input and output sources in simulations as a function of their output
fluxes instead of the (usually considered) input ones. In any
case, this probability function will depend on the input Log N--Log S
and it will be necessary to perform simulations with a number count
distribution as close as possible to the unknown distribution of the
sky sources (an iterative approach would be required). On top of that,
the sensitivity limit of a survey adopting this procedure to correct
for the Eddington bias will be a factor of $\sim 2$ brighter than the
correction of fluxes discussed above. Thus we followed the first
procedure.

\subsection{Final tests and flux limits}

In order to investigate any systematic error in our analysis we
verified how the entire procedure works on different and independent
simulated samples. Thus, we have generated different samples of
sources with different slopes and normalizations of the integral flux
distribution, comparing the input sample with the output of the 
detection procedure, after the correction for the flux
bias and the sky coverage.  Again, to assess the goodness of our
analysis and to derive the flux limit, we perform a maximum
likelihood fit to the differential (unbinned) flux distributions.

\begin{figure*} [hbt]
\begin{tabular}{cc}
{\psfig{figure=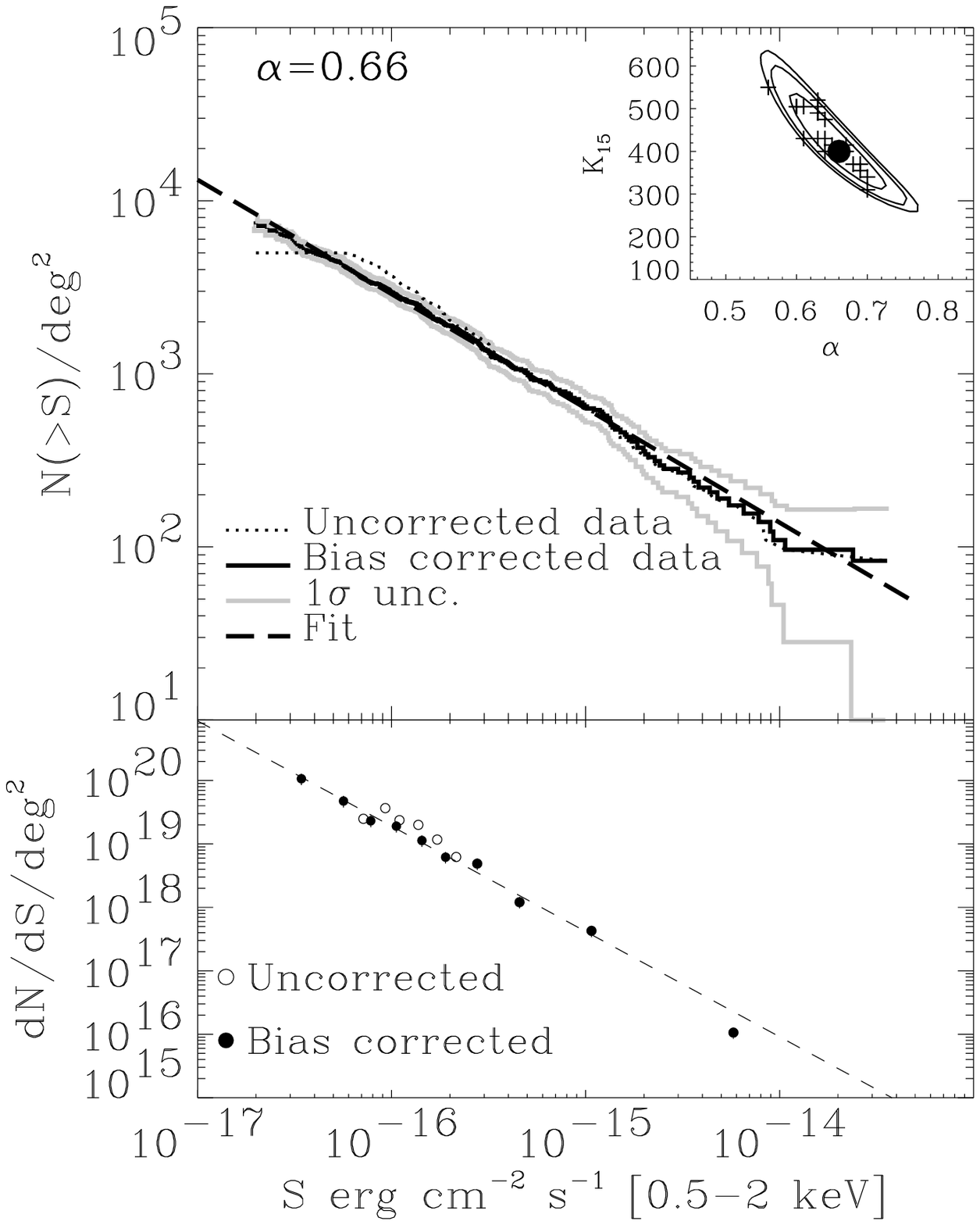,width=7cm}}&{\psfig{figure=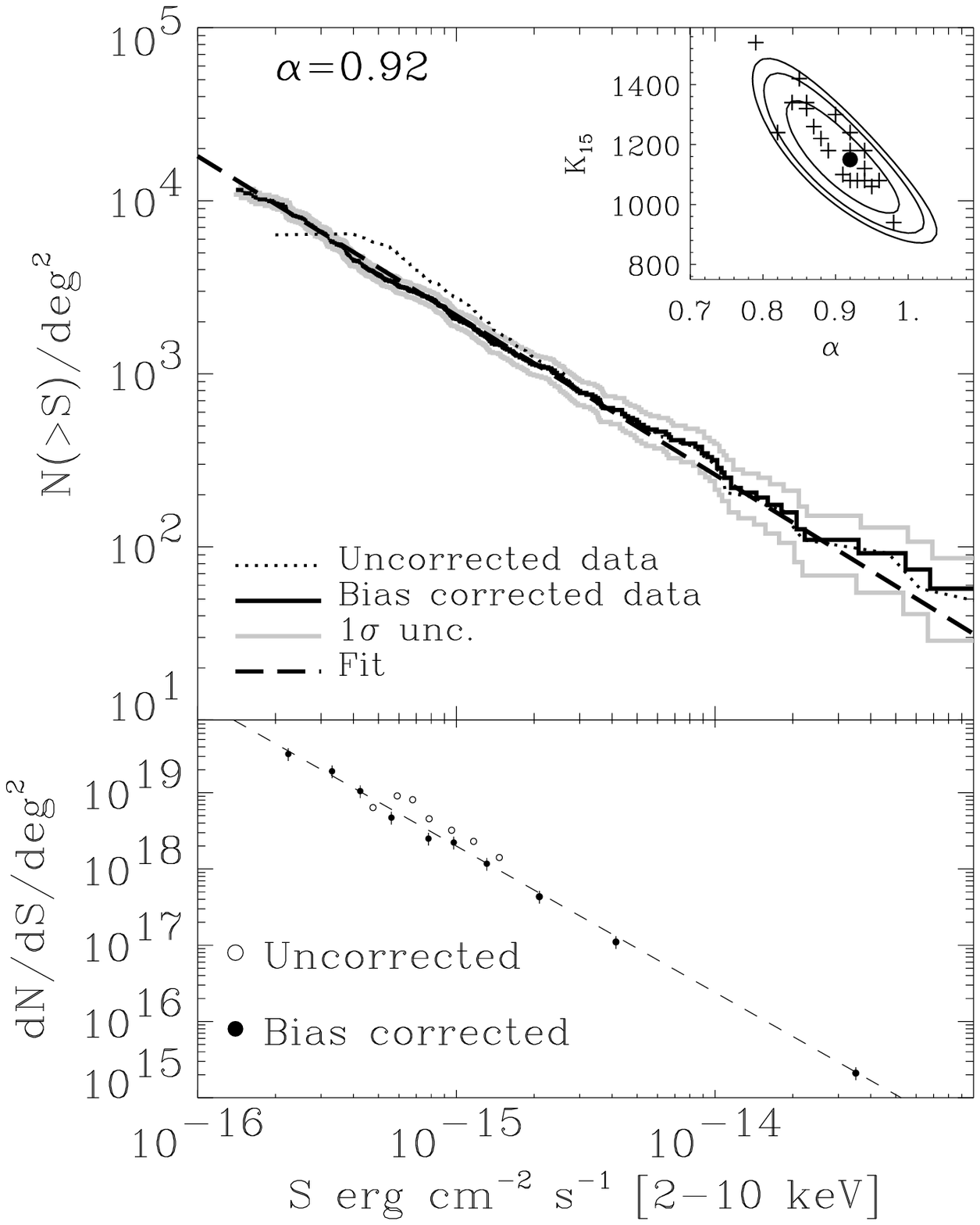,width=7cm}}\\
{\psfig{figure=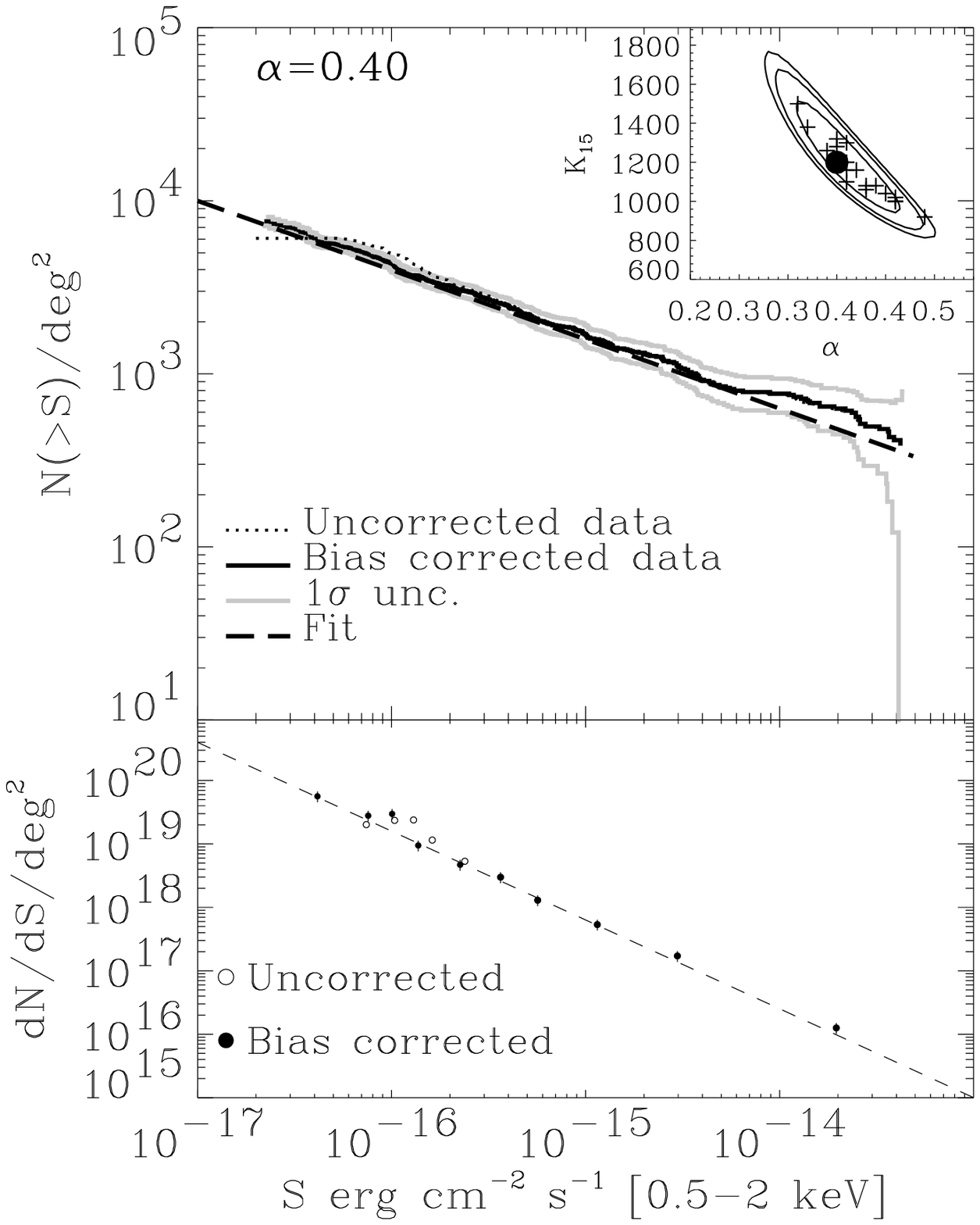,width=7cm}}&{\psfig{figure=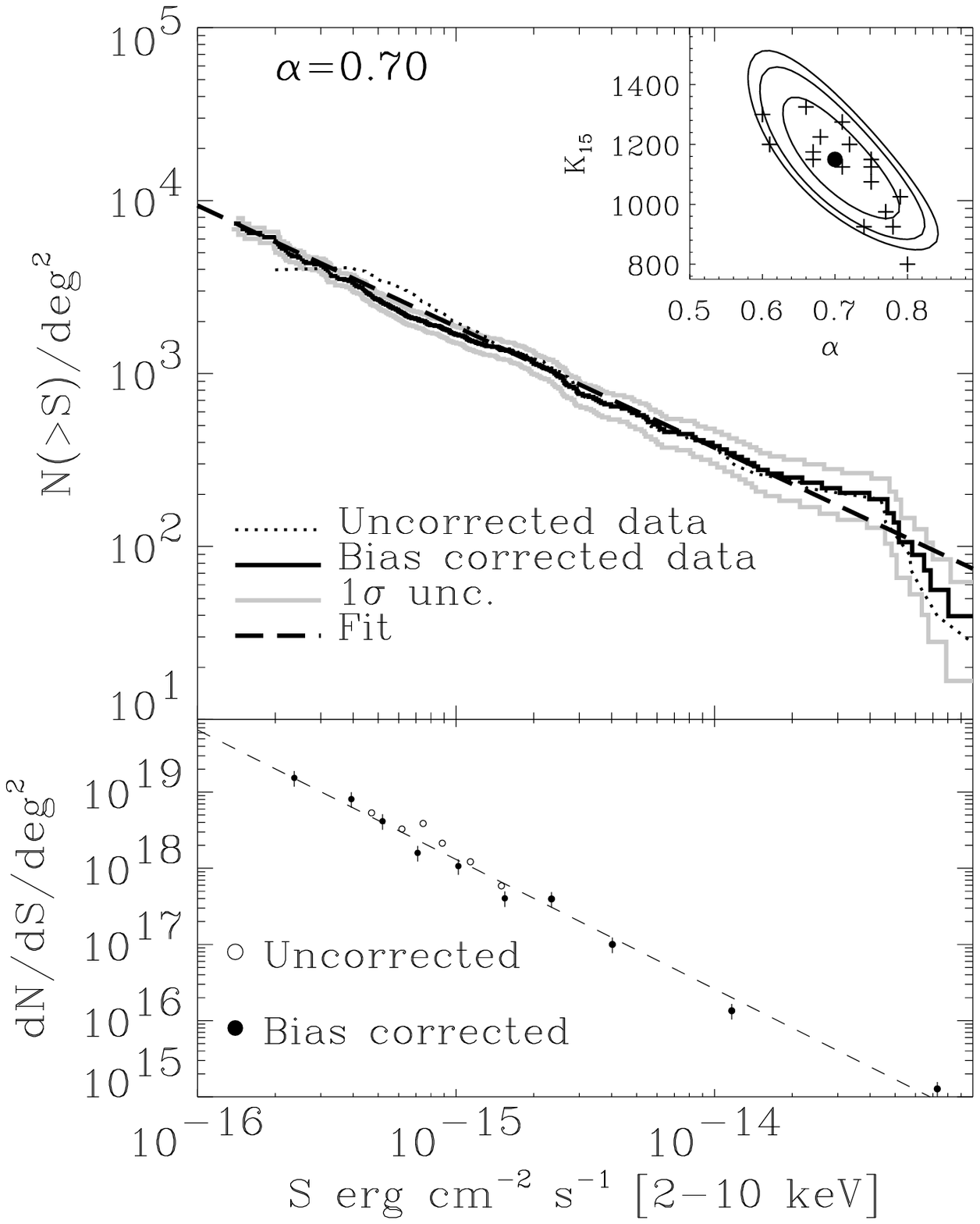,width=7cm}}\\
\end{tabular} 
\caption[]{We report the results of four sets of simulations, two in the soft
and two in the hard band. Each set has different slope and normalization of 
input fluxes and consists of about 15 fields.
In each field about 300 sources are present. In the
main upper panels we report the results relative to one field for each set:
we compare the input integral distribution (dashed line) with the one
recovered after the detection (dotted line) and  after the bias
correction (solid line). In the inserts we plot the results from the
maximum likelihood fits relative to every field of the set (crosses) and the 
expected value (black dot).
In addition we overplot the uncertainties ($68\%,\ 90\%,\ 99\%$ confidence 
level).
In the lower panels the differential distributions are plotted: the data have
been adaptively smoothed to have the same number of sources per bin.}

\label{fig:inout} 
\end{figure*}

In Fig. \ref{fig:inout} we compare the input integral flux
distribution (dashed line) with the output one before (dotted line)
and after (solid line) the correction for the Eddington bias.  As
stated above, without the flux correction for this bias, a reliable flux
limit is about a factor 2 larger. This happens
because all the detected sources with input fluxes in the range 5--10
counts range are detected typically with 10--15 counts.  As a
consequence the slope of the faint tail of the distribution of
uncorrected fluxes is overestimated. This effect depends on the
background level and on the slope of the distribution. It is more 
important in the hard band where the background is higher, 
and where the source distribution has a larger slope.  
As illustrated in Fig. \ref{fig:inout} in all cases
we are able to recover the input distributions with high accuracy and with 
different input slopes down to 5 and 7 counts in the soft and hard 
band, respectively (corresponding to fluxes of $2.4 \times 10^{-17}$ 
and $2.1\times 10^{-16}$ erg s$^{-1}$ cm$^{-2}$).
These numbers refer to the innermost regions and rising to 5, 6 and 8
counts in the other regions (7, 7 and 10 in the hard band).


\section{Log N--Log S distribution}

\begin{figure*} [hbt]
\begin{tabular}{cc}
{\psfig{figure=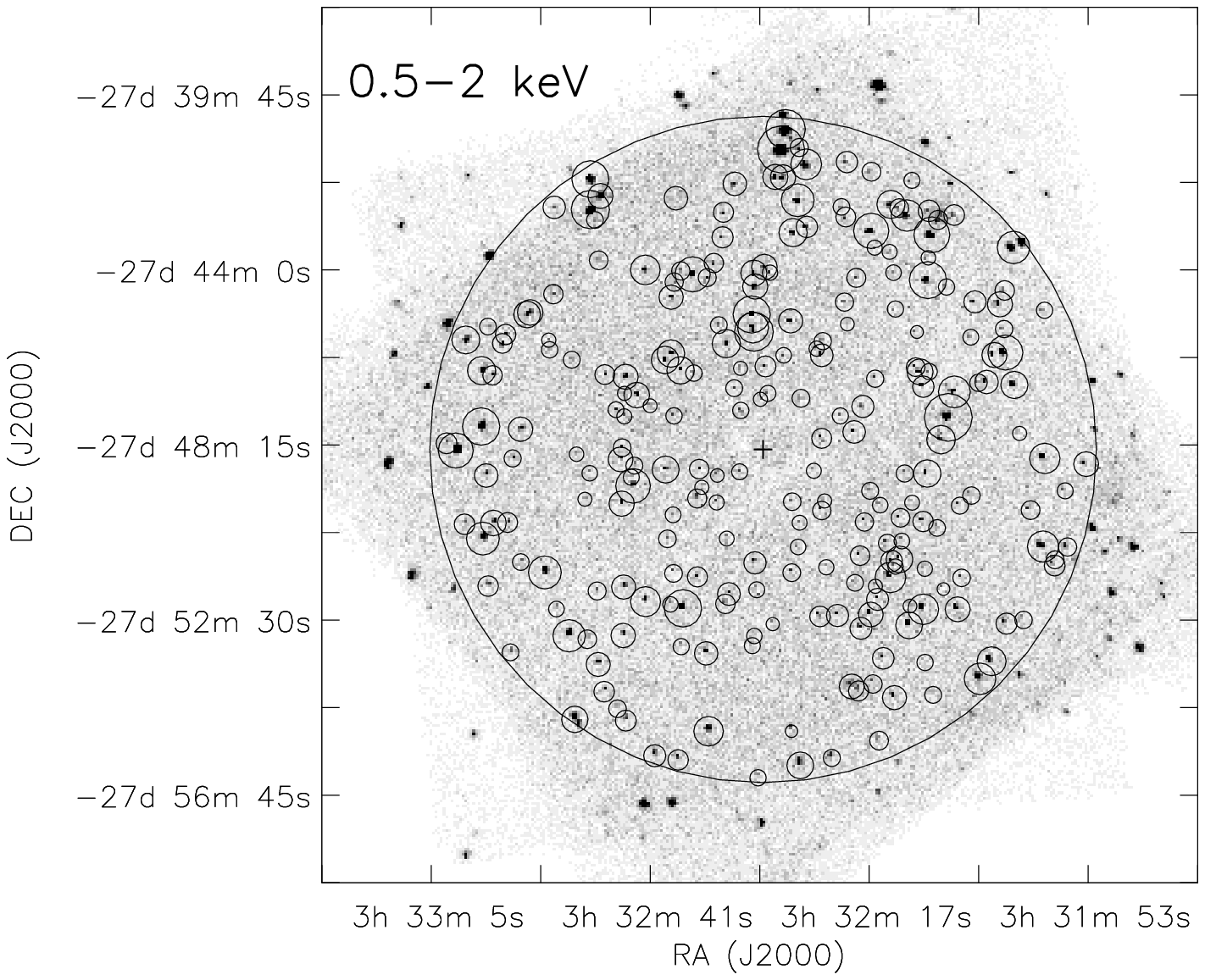,width=8cm}}&{\psfig{figure=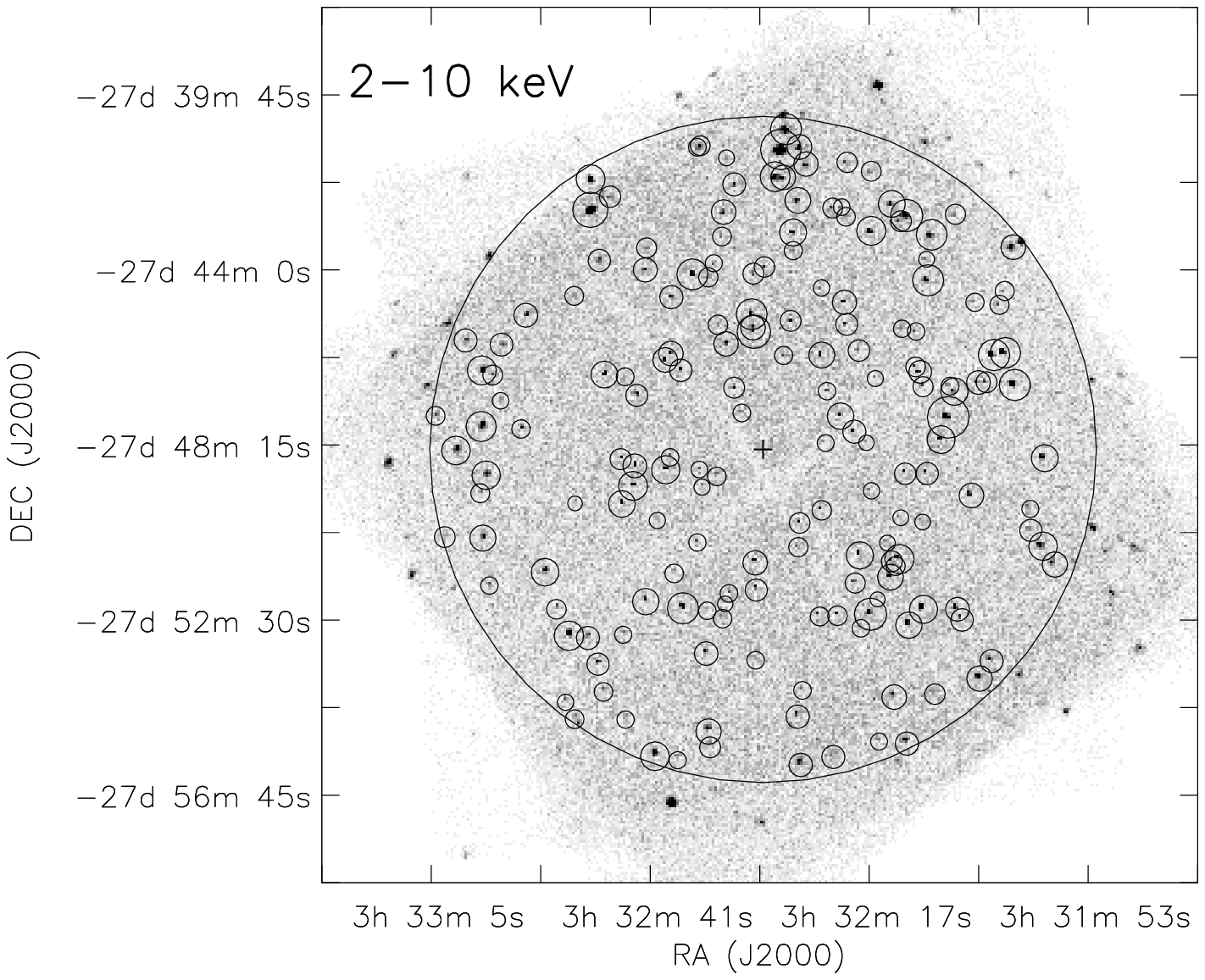,width=8cm}}\\
\end{tabular} 
\caption[]{The detected sources in the soft band (left) and in the hard band
(right). The dimensions of the circles are proportional to the
logarithm of the estimated flux. We restricted our analysis to the
central 8 arcmin radius circle, assuming as center the aim point of the
observation ID581.}
\label{fig:images} 
\end{figure*}

Within the $8'$ radius region in which our analysis 
has been carried out we detected 244 and 177 sources in the soft and hard 
band, respectively (Fig. \ref{fig:images}). These sources are detected
independently in the two images for a total number of 278 sources. 
This is at variance with e.g. Rosati et al. (2001) who made a detection over
the entire 0.5--7 keV energy band and then pick up sources with signal to noise 
ratio larger than 2.1 in the 0.5--2 keV and 2--7 keV energy bands.
Given a typical source spectrum and taking into account 
the different background levels in the two bands this latter approach is
in general more efficient in the source detection, because it uses all
the signal available, but it is not well suitable for our purposes. Infact,
our final aim is drawing the flux distributions in the two bands down to the
faintest limits, where the detection is highly incomplete
and to do this we based on the results of the Monte Carlo simulations:
if we perform the detection and the photometry in the two bands indipendently
we can  simulate all the procedures directly, whereas if we perform the
detection in the full band we have to assume an input spectrum for the sources
to simulate the photometry in the two bands making the evaluation of the sky
coverage at the faint end tricky.

\begin{figure*} [hbt]
\begin{tabular}{cc}
{\psfig{figure=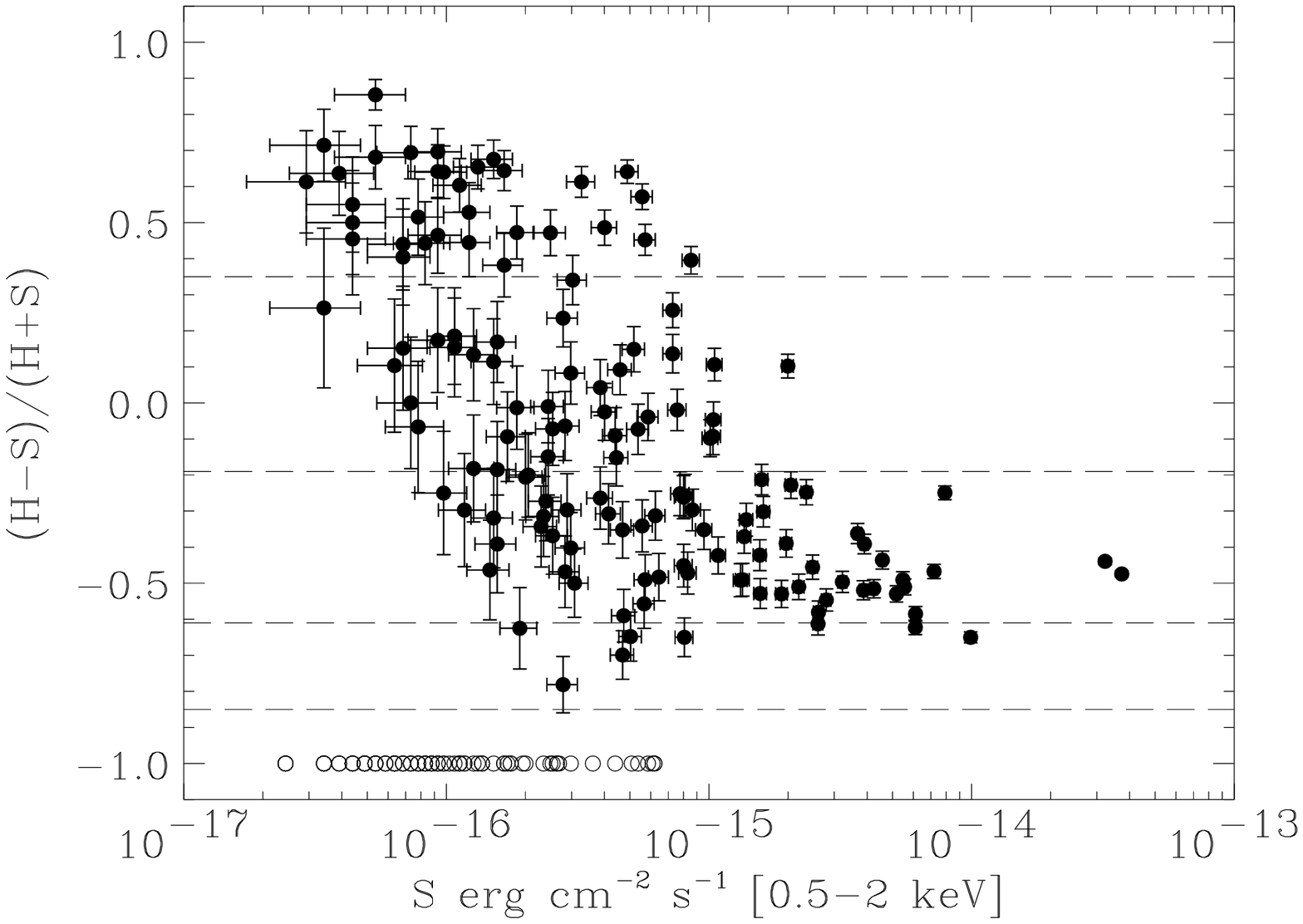,width=12cm}}\\
{\psfig{figure=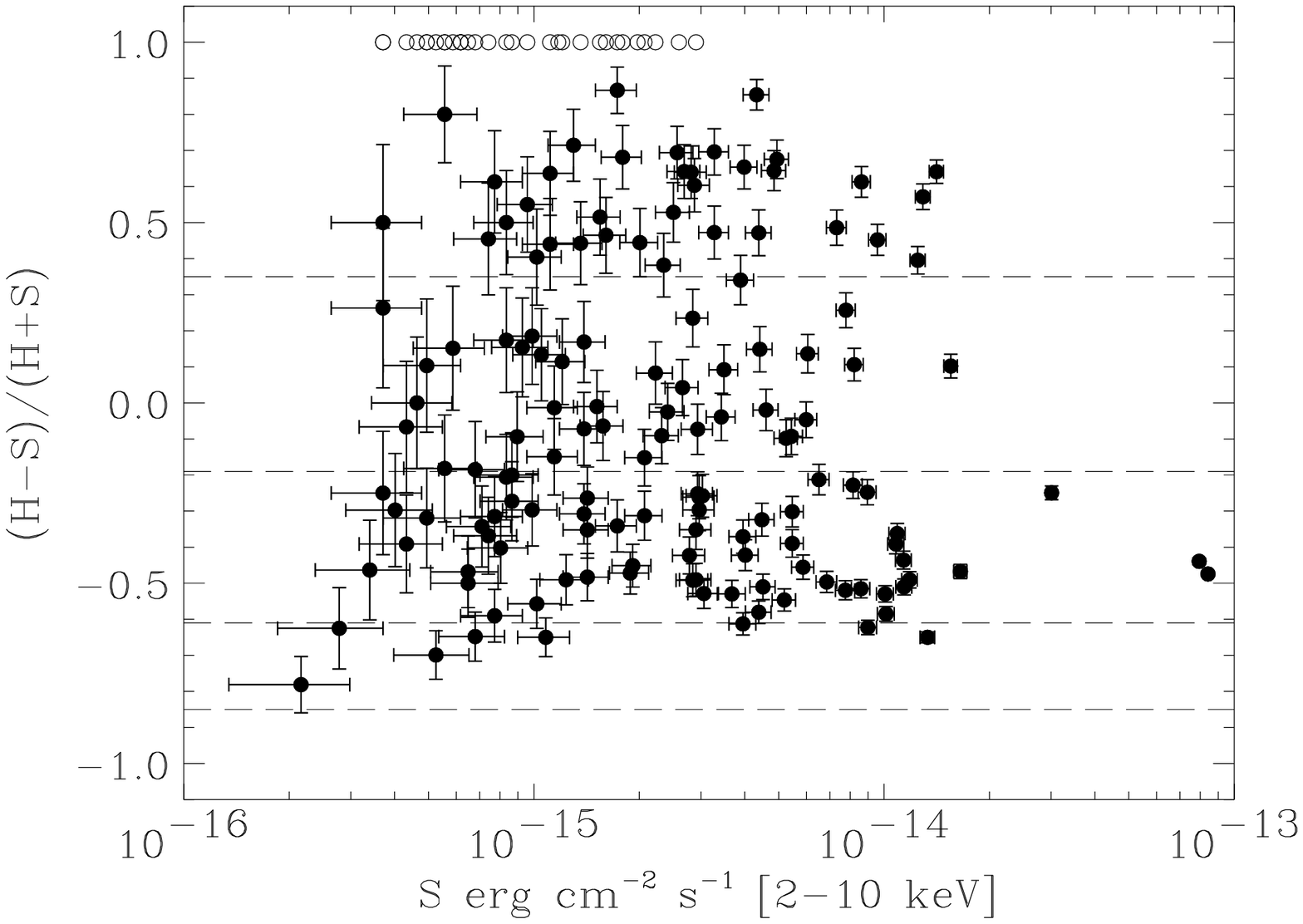,width=12cm}}\\
\end{tabular} 
\caption{{Color diagram of the detected sources versus the soft flux (upper panel)
and hard flux (lower panel). Sources undetected in the hard and soft band are marked with
open circle with a color of --1 and 1, respectively. Dashed lines refer to hardness ratio 
of power law models with photon index 3, 2, 1 and 0 (from bottom to top), respectively.}
\label{fig:hr}}
\end{figure*}

To further characterize the detected sources we computed the hardness
ratio $HR = (H-S)/(H+S)$ where $H$ and $S$ are the net counts in the
hard and the soft band corrected for PSF and vignetting losses,
respectively. There are 34 sources ($\sim 12\%$ of the total number of
detected sources) that are revealed only in the hard band ($HR=1$), and 101
sources ($\sim 36\%$) that are revealed only in the soft band ($HR=-1$). 
As can be seen in Fig. \ref{fig:hr}, there is a progressive hardening of the
detected sources when plotted as a function of the soft flux indicating that
the deeper we go in the soft band, the larger the number of sources
responsible for the XRB at higher energies.  This is particularly
evident at 0.5--2 keV fluxes larger than $\sim 10^{-15}$ erg s$^{-1}$ cm$^{-2}$
for which no source with $HR>0$ are observed (e.g. Tozzi et al. 2001; Brandt
et al. 2001b), 
whereas the large number of sources at low fluxes undetected in the
hard band ($HR=-1$) may weaken the trend. Sources with no soft X--ray
counterparts are likely candidate for type II quasar (Norman et
al. 2001; Stern et al. 2001). The small number of hard
sources not detected in the soft band indicate that even in highly
obscured AGN a sizeable soft emission can still be produced due to,
e.g., scattering, partial covering of the central radiation or from
starburst emission around the AGN (e.g. Turner et al. 1997). This
confirms that also at these extreme flux levels the ratio between soft
and hard X--ray emission is $\sim 1-10\%$. In addition, as shown in 
Barger et al. (2001), the effects of obscuration get significantly reduced 
at $z\sim 2$, since we observe in the soft band the harder 1.5--6 keV
rest frame interval.

\begin{figure*}[hbt]
\begin{tabular}{cc}
{\psfig{figure=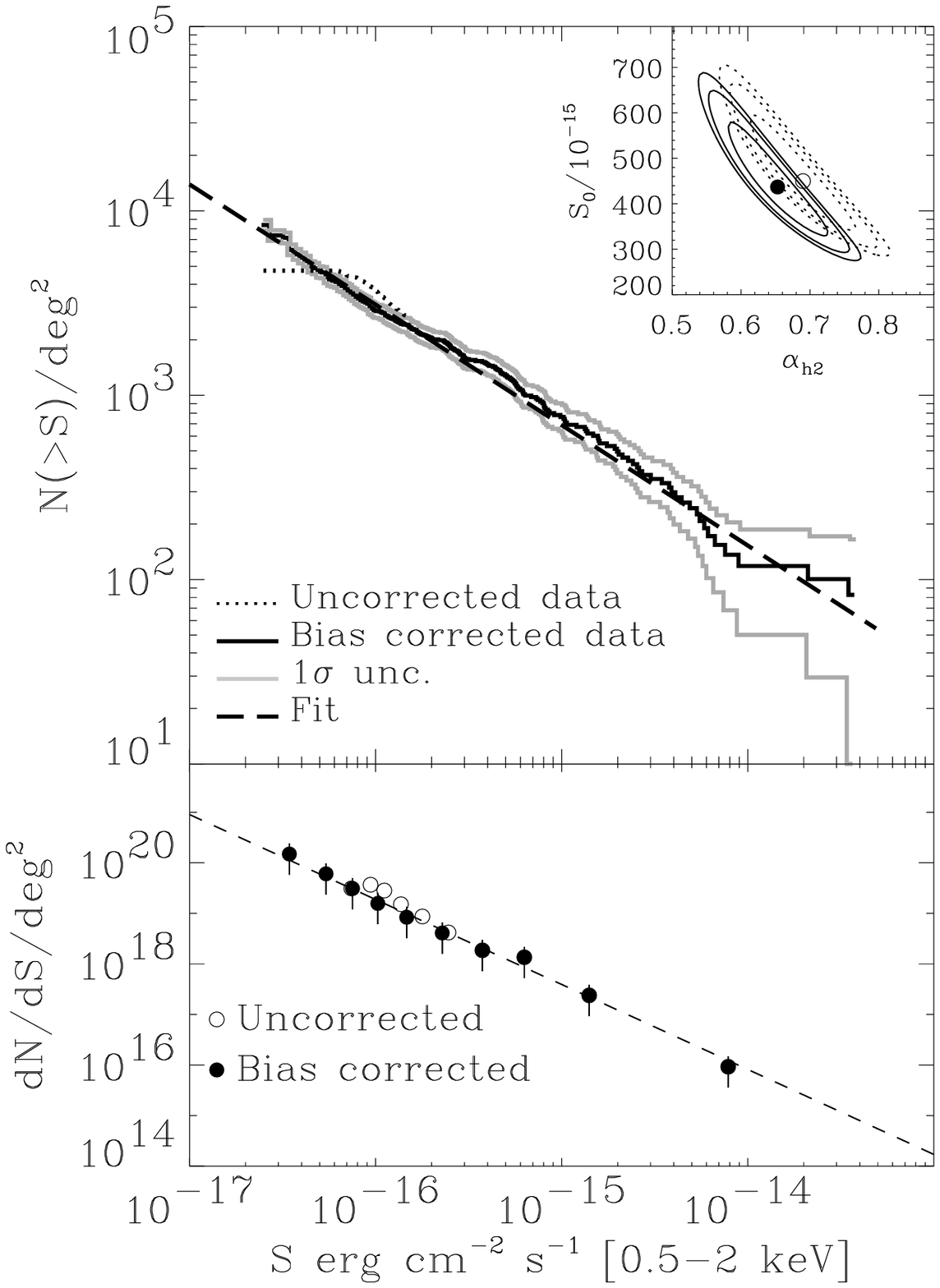,width=8cm}}&{\psfig{figure=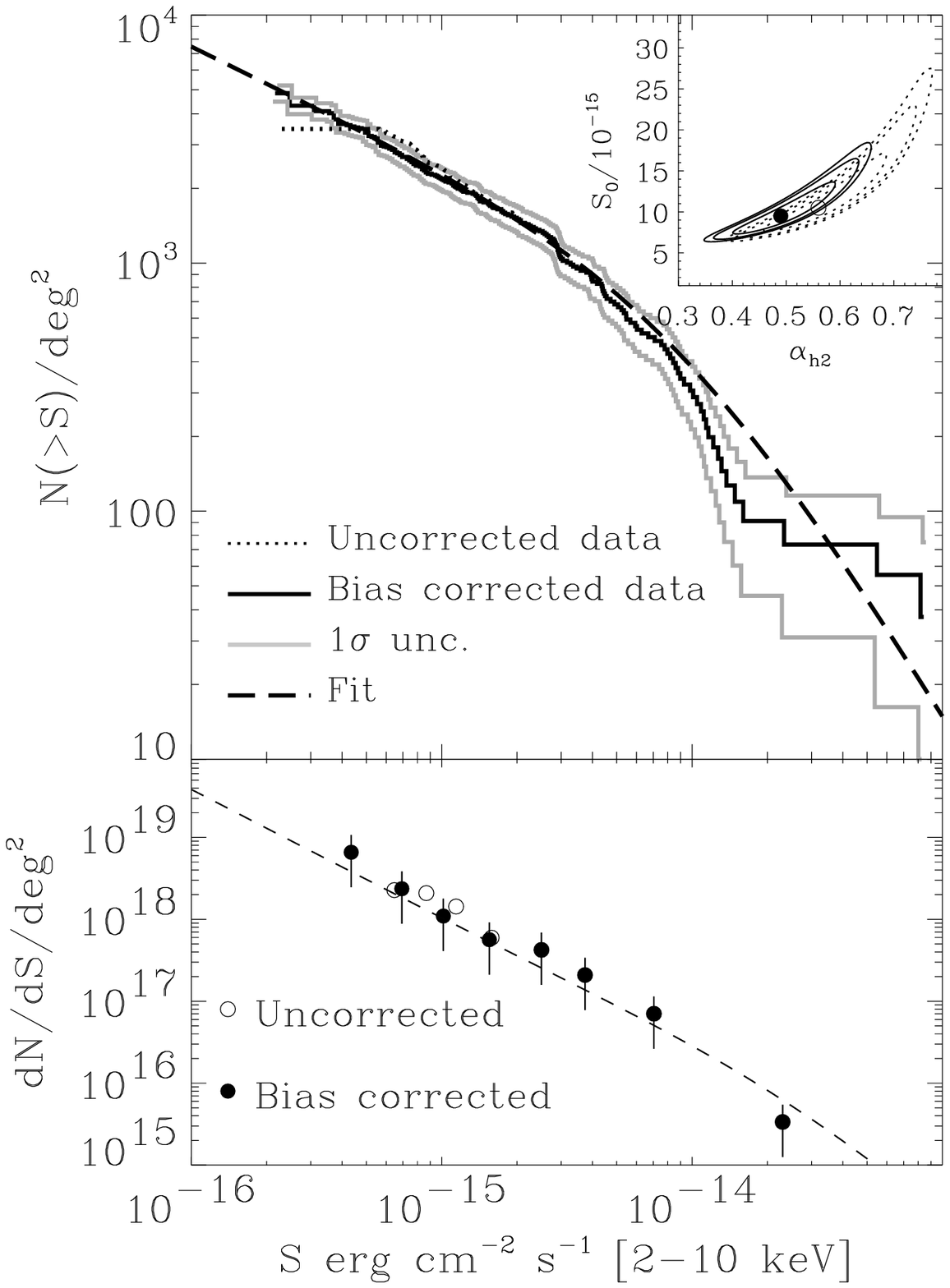,width=8cm}}\\
\end{tabular} 
\caption[]{Left panel: Soft band Log N--Log S from inner $8'$ radius of 
the 1~Ms observation of the CDFS (thick solid line) in the range
$2\times10^{-17}-3\times10^{-14}$ erg s$^{-1}$ cm$^{-2}$for an
average $\Gamma=1.4$ power--law spectrum. The gray thick line shows
the $1\sigma$ confidence region taking into account the number
statistics and the flux conversion uncertainties. The thick dashed
solid line shows the best maximum likelihood fit. The dotted line represents 
the Log N--Log S distribution {\it without} the correction for the Eddington
bias. In the insert we plot the best value and the uncertainties
for the corrected fluxes (filled circle and continuous line)
and the uncorrected ones (empty circle and dashed lines).\\ 
Right panel: Hard band Log N--Log S from inner $8'$ radius of
the 1~Ms observation of the CDFS (thick solid line) in the range
$2\times10^{-16}-2\times10^{-14}$ erg s$^{-1}$ cm$^{-2}$. Lines are as
in the right panel. 
In the lower panels the differential distribution of corrected fluxes 
(filled circles) and uncorrected fluxes (open circles) are plotted with the 
fit. The data have been adaptively binned to have the same number of sources
per bin}
\label{fig:lnlss}
\end{figure*}

In order to characterize the flux distribution of sources detected in the soft 
band we performed a maximum likelihood fit with a power--law distribution. 
We obtain as a best fit
\begin{equation} 
N(>S) = 438\, \Bigl( {{S}\over{2\times 10^{-15}}}\Bigr) ^{-0.65} \ {\rm cgs}.
\end{equation}
The error at $68\%$ confidence level for the normalization is
$K_{15}=438^{+82}_{-68}$ ($K_{15}$ means that fluxes are measured in
units of $2\times 10^{-15}$ erg s$^{-1}$ cm$^{-2}$), while for the
slope is $\alpha_{\rm s}=0.65\pm0.05$ (see also Fig. \ref{fig:lnlss}).
This value is slightly different from the one reported in Campana et
al. (2001) by less than $1\,\sigma$. This is mainly due to the improvements 
in the total background map calculations.
Maximum likelihood analysis provide an accurate description of the errors
only in the case of good fits. To assure this we adaptively binned the 
source distribution in order to
contain 10 sources per bin. The fit of the binned data with the
same power law gives a good fit by means of a $\chi^2$ test
(null hypothesis probability $\sim 10\%$).

In order to test the influence of the Eddington bias correction in the
fit of the slope and normalization of the Log N--Log S, we fit with the
same maximum likelihood routine the uncorrected number counts.
As explained above, a further implication of the Eddington bias correction 
is that the detection can be performed at fainter fluxes.
Our soft band limiting fluxes are, in fact, $F_{\rm lim}=4.9\times10^{-17}$
and $F_{\rm lim}=2.4\times10^{-17}$ erg s$^{-1}$ cm$^{-2}$ before and after
the correction, respectively.
We find that the normalization does not change significantly
($K_{15}=420^{+80}_{-68}$) while the slope is only marginally
consistent with what measured after the bias correction ($\alpha_{\rm
s}=0.72\pm0.05$; see also Fig. \ref{fig:lnlss}).

Our bias corrected normalization and slope, are in good agreement with
earlier results obtained with ROSAT (Hasinger et al. 1998), Chandra (Tozzi
et al. 2001) and XMM-Newton (Hasinger et al. 2001; Baldi et al. 2001) up to
their faintest flux levels ($>10^{-16}$ erg s$^{-1}$ cm$^{-2}$).  

The total extragalactic background in the soft band is usually evaluated in
the 1--2 keV band to limit contribution from our Galaxy. The absolute value of 
the XRB is still a matter of debate. The 1--2 keV XRB
estimates are clustered around two different values: Hasinger et
al. (1993) and Gendreau et al. (1995) obtained a value of $3.76\times
10^{-12}$ erg s$^{-1}$ cm$^{-2}$ deg$^{-2}$. This is in contrast with the
estimate derived by Chen et al. (1997; see also Hasinger 1996) based on a
joint ROSAT/ASCA fit yielding a flux of $4.42\times 10^{-12}$ erg s$^{-1}$
cm$^{-2}$ deg$^{-2}$ (i.e. $\sim 20\%$ higher). An intermediate value has
recently been obtained by Kuntz, Snowden \& Mushotzky (2001) using ROSAT PSPC
observations with $3.99\times 10^{-12}$ erg s$^{-1}$ cm$^{-2}$ deg$^{-2}$ and
Garmire et al. (2002, in preparation) using Chandra $4.19\times 10^{-12}$ erg
s$^{-1}$ cm$^{-2}$ deg$^{-2}$. 
Based on our Log N--Log S distribution we can resolve in point sources a flux
of $5.58\times 10^{-13}$ erg s$^{-1}$ cm$^{-2}$ deg$^{-2}$ for sources fainter
than $10^{-15}$ erg s$^{-1}$ cm$^{-2}$. Summing to this the contribution from
the Lockman hole (Hasinger et al. 1998) we obtain a resolved flux of
$3.57\times 10^{-12}$ erg s$^{-1}$ cm$^{-2}$ deg$^{-2}$. 
As pointed out by Rosati et al. (2001) we have also to include the
contribution from cluster of galaxies recognized to be at a level of $6\%$ in
the 1--2 keV energy band (Rosati et al. 1998).
In conclusion, in the case of the lower background estimate we are able to
fully resolve it as  arising from point sources, taking the higher value we
can resolve $87\%$. 

The slope of the faint end counts converges slower than logarithmically and
therefore the faint sources have a very small contribution to the XRB.
Extending to lower fluxes the source distribution we can add just 
a $4\%$ more. This may leave space for a $\lsim 9\%$ truly diffuse emission
produced by the warm/hot diffuse intergalactic medium which is estimated to
contribute at a $5-15\%$ level (Phillips et al. 2001).

A final caveat concerns the total flux of the point sources which
suffers from several uncertainties: source flux conversion, variation of the
power law index from source to source, cosmic variance (the CDFS
is depleted of bright sources), Log N--Log S power law index and
normalization uncertainties. We estimate an error at a level of $7\%$.

In the case of the hard band,
a fit with a single power-law source distribution does not provide a
good description of the data (binning the data as before, we obtain a
probability of the null hypothesis $\sim 0.02\%$). 
Moreover due to the lack of bright sources in the CDFS,
our data cannot be matched directly with earlier result. So we normalized the
distribution of our 177 sources, see Fig. \ref{fig:lnlss}) to the ASCA data
(della Ceca et al. 2000). The limiting flux is $2.1\times 10^{-16}$ erg s$^{-1}$
cm$^{-2}$ and raises to $F_{\rm lim}=3.7\times10^{-16}$ erg
s$^{-1}$ cm$^{-2}$ if the Eddington bias correction is not applied.

Thus, we turn to a smoothly joined power--law model:
\begin{equation} 
N(>S) = 1.2\times 10^4\, \Bigl[ 
{{(2\times 10^{-15})^{\alpha_{\rm h1}}} \over
{S^{\alpha_{\rm h1}}+S_0^{\alpha_{\rm h1}-\alpha_{\rm h2}}\,
S^{\alpha_{\rm h2}}}} \Bigr] \ {\rm cgs},
\end{equation}

\noindent fixing the bright flux slope and the normalization to the ASCA
value, i.e.  $\alpha_{h1}=1.67$ (della Ceca et al. 2000). The faint
end slope and the break flux $S_0$ are fitted to the data. The maximum
likelihood fit yields $\alpha_{\rm h2}=0.49\pm0.05$ and
$S_0=9.6^{+1.2}_{-1.0}\times 10^{-15}$ erg s$^{-1}$ cm$^{-2}$ (all at
$68\%$ confidence level for a single parameter; see Fig. \ref{fig:lnlss}).
Without correcting for the Eddington bias, the break flux slightly
changes $S_0=10.5^{+1.2}_{-1.0}\times 10^{-15}$ and the fitted slope is
$\alpha_{\rm h2}=0.56\pm0.05$. 
The slope value is slightly flatter (within $1\,\sigma$) than what
found in Campana et al. (2001).

The problem of estimating the background level in the hard 2--10 keV band is
even more serious than in the soft band. Also in this case different estimates 
exist but the range of variation is as high as $50\%$. 
The limiting values are on the lower side the ones from UHURU and HEAO-1 
$1.6 \times 10^{-11}$ erg s$^{-1}$ cm$^{-2}$ deg$^{-2}$ (Marshall et al. 1980; 
see also Ueda et al. 1999 based on ASCA)
and on the higher side the (more recent) ones from BeppoSAX and ASCA 
$2.4 \times 10^{-11}$ erg s$^{-1}$ cm$^{-2}$ deg$^{-2}$ (Chiappetti
et al. 1998; Vecchi et al.  1999; Ishisaki et al. 2000; Perri \& Giommi 2000),
in agreement with the old Wisconsin measurements (McCammon et al. 1983).

With our number counts distribution, we are able to account for 
$1.99\times 10^{-11}$ erg s$^{-1}$ cm$^{-2}$ deg$^{-2}$. This value is higher
than the UHURU and HEAO-1 estimate by $24\%$ likely ruling out it.
With respect to the higher 2--10 keV background estimate our source
distribution can account for $83\%$ of the background resolved in point
sources. Also in this case the faint end tail converges slower than
logarithmically and we can comprise up to $87\%$ of the hard background
(see also Campana et al. 2001).
A new population of sources, possibly contributing also at higher energies, is 
therefore needed if one wants to fully resolve the background into point sources.

In this case, the error on the background flux resolved into sources is 
dominated by the uncertainty in the flux conversion factor, resulting in a
$10\%$ error. 


\section{Conclusions}

Based on the experience developed in the ROSAT HRI data analysis
(Lazzati et al. 1999; Campana et al. 1999) we build up a procedure for
the automatic detection and characterization of sources in Chandra
ACIS fields. The main difference between our wavelet algorithm
(BMW-Chandra) and the standard CIAO {\tt wavdetect} (Freeman et
al. 2001) is the use of the multi-resolution technique. As explained,
this makes the BMW-Chandra algorithm less flexible but faster and therefore
more suitable for the analysis of large data sets and for Monte Carlo
tests. The Chandra deep fields are, of course, the most intriguing
fields to probe the algorithm. Here we have described the procedures we
used to simulate and analyze in details the CDFS. The simulations we
performed allow us to safely conclude that our detection procedures
are robust in terms of number of spurious sources, positional
accuracy, flux estimate down to the quoted limits. In this respect the
Eddington bias correction is especially important, allowing us to
approach the theoretical detection limit of 3 photons (Damiani et
al. 1997b).

Scientific results have been previously discussed in Campana et al. (2001),  
here we improve and complement them. We independently detect 244 and 177
sources in the soft and hard band, respectively for a total number of 278
sources. Source distributions can be safely recovered down to fluxes
$2.4\times 10^{-17}$ and $2.1\times 10^{-16}$ erg s$^{-1}$ cm$^{-2}$ in the
soft and hard energy band, respectively. These are a factor of $\gsim 2$
deeper than current estimates. Note that being still in the photon detection
limit, an improvement by a factor of $\sim 2$ in the Log N--Log S
reconstruction corresponds to an effective doubling of the observing time. 

There is general consensus that the main constituents of the soft 1--2 keV XRB 
can be ascribed to broad line AGN (i.e. Seyfert 1 galaxies; Hasinger et
al. 1998; Schmidt et al. 1998). At faint fluxes ($\lsim 10^{-15}$ erg s$^{-1}$
cm$^{-2}$) nearby ($z\lsim 0.6$) optically-normal (possibly low luminosity
AGN) galaxies are also being detected as soft sources (Fiore et al. 2000;
Barger et al. 2001; Tozzi et al. 2001; Brandt et al. 2001a; Schreier et
al. 2001; Koekemoer et al. 2001). Our analysis of the CDFS 
extends the source flux distribution down the faintest level of $2.4\times
10^{-17}$ erg s$^{-1}$ cm$^{-2}$. We found that even at these low fluxes the
Log N--Log S distribution can still be represented by the extrapolation from
higher fluxes without upward trends. Including the contribution from cluster
of galaxies (Rosati et al. 1998), we are able to resolve in point sources
$>90\%$ of the soft XRB, the exact value depending on the sky level itself.
Our results are also in agreement with the recent fluctuation analysis on the
HDFN down to $7\times 10^{-18}$ erg s$^{-1}$ cm$^{-2}$ (Miyaji \& Griffiths
2002). 
 
In the hard band the XRB is made by the superposition of absorbed and
unabsorbed sources (Setti \& Woltjer 1989).
The steeper power law index observed in bright AGN implies
that a population of faint hard sources is present, likely affected by a
considerable absorption. Studies with Chandra and XMM-Newton are now discovering
these sources identifiable with Type II AGN unrelated to the morphological
type (Schreier et al. 2001; Koekemoer et al. 2001). We note that only a small
fraction of hard source are not detected in the soft band, likely indicating
that at least $1-10\%$ of the flux is in any case emitted in the soft band.
Our analysis lead to the extension of the Log N--Log S source flux
distribution down to $2.1\times 10^{-16}$ erg s$^{-1}$ cm$^{-2}$.
At this level we can rule out the UHURU/HEAO-1 estimate and account for 
$87\%$ of the BeppoSAX/ASCA estimate. The fluctuation analysis by Miyaji \&
Griffiths (2002) does not add much stopping at $1\times 10^{-16}$ erg s$^{-1}$
cm$^{-2}$ but with a large error. 

The analyzed field of view is small and cosmic variance likely plays a role. 
We plan to carry out the same analysis of the Chandra HDFN in order to confirm
the present results and to perform a joint analysis with other shallower and 
more extended surveys in order to cover with good statistics the bright flux
ends (Moretti et al., in preparation).

Chandra with its superb angular resolution opens the possibility to extend 
at even lower fluxes the source distribution. In fact, at the present level 
sources are still photon limited. This will allow to deepen our knowledge of
the distant universe, allowing to probe the accretion power over the history 
of the X--ray universe, and its implications for structure formation and the 
epoch of reionization. 

\begin{acknowledgements}
We thank the continuous support of the Chandra Help Desk and the CIAO team for
the organization of Chandra/CIAO workshop. We also thank M. Markevitch and the 
anonymous referee for their suggestions.
This work was supported through CNAA, Co-fin, ASI grants and Funds for Young
Researchers of the Universit\`a degli studi di Milano. 
\end{acknowledgements}

\clearpage

\begin{table*}[htb]
\begin{center}
\caption[]{Main characteristics of the 11 single exposures.}
\footnotesize{
\begin{tabular}{|l|cccccccccc|}
\hline
ID     &R.A.    & Dec.   &   Roll    &Soft.Exp. & Hard.Exp.&Nom.Expos.&$\Delta \alpha$&$\Delta \delta$& $\sigma_{\Delta\alpha}$ &$\sigma_{\Delta \delta}$\\
       &(J2000) &(J2000) &   Angle   & (ks)     & (ks)     & (ks)     & (pixel)       & (pixel)       & (pixel)                 & (pixel) \\
\hline
0581& 53.122&  --27.806&   47.28&    15.594&  15.311 &   18.713&\    0.0&\    0.0&  0.00& 0.00\\ 
0441& 53.111&  --27.804&  166.73&    55.967&  54.967 &   55.967&    72.8&\    7.6&  0.51& 0.65\\ 
0582& 53.111&  --27.804&  162.92&   130.486& 128.595 &  130.486&    70.9&\    8.6&  0.77& 0.65\\ 
1431& 53.122&  --27.806&   47.28&   110.105& 107.560 &  123.008&\    1.2&\  --0.7&  0.64& 0.61\\ 
1672& 53.120&  --27.813&  326.90&    95.138&  95.138 &   95.138&    20.2&  --51.1&  0.79& 0.86\\ 
2239& 53.117&  --27.811&  319.20&   123.999& 129.732 &  130.738&    38.9&  --39.7&  0.77& 0.81\\ 
2312& 53.118&  --27.811&  329.91&   123.686& 123.686 &  123.686&    31.2&  --35.4&  0.63& 0.72\\ 
2313& 53.117&  --27.811&  319.20&   130.389& 130.389 &  130.389&    37.7&  --39.0&  0.62& 0.59\\ 
2405& 53.120&  --27.812&  331.80&    58.215&  55.725 &   59.635&    16.4&  --50.1&  0.70& 0.67\\ 
2406& 53.118&  --27.810&  332.18&    29.686&  28.696 &   29.686&    30.7&  --34.3&  0.91& 0.59\\ 
2409& 53.117&  --27.811&  319.20&    68.982&  68.982 &   68.982&    38.8&  --39.1&  0.92& 1.02\\ 
\hline 
\end{tabular}}
\label{tab:subexp}  
\end{center}
\noindent Pixels are at nominal rebin (i.e. $0.49''$).
\end{table*}

\begin{thebibliography}{99}

\bibitem[]{796} 
Baganoff, F. K. 1999, ACIS Memo 162,\\
{\tt
http://asc.harvard.edu/cal/Links/Acis/acis/Cal\_prods/bkgrnd/current/index.html}.

\bibitem[]{801} 
Baldi, A., Molendi, S., Comastri, A., Fiore, F., Matt, G., Vignali, 
C. 2001, ApJ in press (astro-ph/0108514)

\bibitem[]{805} 
Barger, A. J., Cowie, L. L., Mushotzky, R. F., Richards, E. A. 
2001, AJ, 121, 662

\bibitem[]{809} 
Bautz, M., et al. 1998, SPIE, 3444, 210

\bibitem[]{812} 
Brandt, W. N., et al. 2001a, AJ, 122, 1

\bibitem[]{815} 
Brandt, W. N., et al. 2001b, AJ, in press (astro-ph/0108404)

\bibitem[]{818}
Bijaoui A., Giudicelli, M. 1991, Exp Astron., 1, 347 

\bibitem[]{821}
Campana, S., et al. 1999, ApJ, 524, 423

\bibitem[]{824}
Campana, S., Moretti, A., Lazzati, D.,  Tagliaferri, G. 2001, ApJ 560 L19

\bibitem[]{827} 
Chiappetti, L., Cusumano, G., del Sordo, S.,
Maccarone, M.C., Mineo, T., \&  Molendi, S. 1998, in {\it The Active X-ray Sky:
Results from BeppoSAX and RXTE.} eds. L. Scarsi, H. Bradt, P. Giommi, \&
F. Fiore. (Elsevier, Amsterdam), 610


\bibitem[]{837} 
Damiani, F., Maggio, A., Micela, G.,  Sciortino, S. 1997a, ApJ, 483, 350 

\bibitem[]{840} 
Damiani, F., Maggio, A., Micela, G.,  Sciortino, S. 1997b, ApJ, 483, 370 

\bibitem[]{843} 
della Ceca, R., Braito, V., Cagnoni, I., Maccacaro, T. 2000, Proceedings of
the Conference SAIt 2000, Mem. SAIt in press

\bibitem[]{847} 
Fiore, F., et al. 2000, New Astr., 5, 143

\bibitem[]{850} 
Freeman, P. E., et al. 2001, ApJ, in press (astro-ph/0108429)

\bibitem[]{853} 
Garmire, G. P., et al. 2002, ApJ submitted 

\bibitem[]{856} 
Gendreau, K.C., et al. 1995, PASJ, 47, L5

\bibitem[]{859} 
Giacconi, R., et al. 2001, ApJ, 551, 624


\bibitem[]{865} 
Grebenev, S. A., Forman, W., Jones, C., Murray, S. 
1995, ApJ, 445, 607 

\bibitem[]{869}
Hasinger, G., et al. 1993, A\&A, 275, 1

\bibitem[]{872}
Hasinger, G., et al. 1998, A\&A, 329, 482

\bibitem[]{875} 
Hasinger, G., et al. 2001, A\&A, 365, L45

\bibitem[]{878} 
Hornschemeier, A. E., et al. 2000, ApJ, 541, 49

\bibitem[]{881} 
Koekemoer, A. N., et al. 2001, ApJ in press (astro-ph/0110385)

\bibitem[]{884} 
Kuntz, K. D., Snowden S. L., Mushotzky R. F. 2001, ApJ, 548, L119

\bibitem[]{887} 
Ishisaki, Y., et al. 2000, in {\it Broad
Band X-ray Spectra of Cosmic Sources -- COSPAR }, eds. K. Makishima, L.
Piro, \& T. Takahashi (Pergamon Press)

\bibitem[]{892} 
Lazzati, D., Campana, S., Rosati, P., Chincarini, G., 
Giacconi, R. 1998, A\&A, 331, 41 

\bibitem[]{896}
Lazzati, D., et al. 1999, ApJ, 524, 414

\bibitem[]{899} 
Marshall, F. E., et al. 1980, ApJ, 235, 4

\bibitem[]{902} 
McCammon, D., Burrows, D. N., Sanders,  W. T., Kraushaar,
W. L. 1983, ApJ, 269, 107

\bibitem[]{906} 
Miyaji, T., Griffiths, R. E. 2002, ApJ in press (astro-ph/0111393)

\bibitem[]{909} 
Mushotzky, R. F., Cowie, L. L., Barger, A. J.,  Arnaud, K. A. 
2000, Nat, 404, 459

\bibitem[]{913} 
Norman, C., et al. 2001, ApJ, in press (astro-ph/0103198)

\bibitem[]{916} 
Panzera, M. R., et al. 2002, to be submitted to A\&A

\bibitem[]{919} 
Perri, M., Giommi, P. 2000, A\&A, 362, L57

\bibitem[]{922} 
Pislar, V., Durret, F., Gerbal, D., Lima Neto, G. B., Slezak, E. 1997, 
A\&A, 322, 53

\bibitem[]{926} 
Phillips, L. A., Ostriker, J. P., Cen, R. 2001, ApJ, 554, L9

\bibitem[]{929} 
Rosati, P., della Ceca, R., Burg, R., Norman, C., Giacconi, 
R. 1995, ApJ, 445, L11

\bibitem[]{933} 
Rosati, P., della Ceca, R., Norman, C., Giacconi, R. 1998, 
ApJ, 492, L21 

\bibitem[]{937} 
Rosati, P., et al. 2001, ApJ in press

\bibitem[]{940} 
Schreier, E. J., et al. 2001, ApJ, 560, 127

\bibitem[]{943} 
Schmidt, M., et al. 1998, A\&A, 329, 495

\bibitem[]{946} 
Setti, G., Woltjer, L. 1989, A\&A, 224, L21

\bibitem[]{949} 
Slezak, E., Durret, F., Gerbal, D. 1994, AJ, 108, 1996

\bibitem[]{952}
Stern, D., et al. 2001, ApJ in press (astro-ph/0111513) 

\bibitem[]{955}
Tozzi, P., et al. 2001, ApJ, 562, 42

\bibitem[]{958} 
Ueda, Y., et al. 1999,  ApJ, 518, 656

\bibitem[]{961} 
Vecchi, A., Molendi, S., Guainazzi, M., Fiore, F., Parmar, A. N. 1999, A\&A,
349, L73

\bibitem[]{965} 
Vikhlinin, A., Forman, W., Jones, C., Murray, S. 1995, ApJ 451, 542

\bibitem[]{968} 
Vikhlinin, A., McNamara, B. R., Forman, W., Jones, C., 
Quintana, H., Hornstrup, A. 1998, ApJ, 502, 558 

\bibitem[]{972} 
Weisskopf, M. C., Tananbaum, H. D., Van Speybroeck, L. P.,
O'Dell, S. L. 2000, SPIE, 4012, 2

\end{thebibliography}
\end{document}